\documentclass[12pt,preprint]{aastex}

 \newcommand{\msun}{M$_\odot$}
 \newcommand{\kms}{km s$^{-1}$}
 \newcommand{\HI}{\ion{H}{1}}
 \newcommand{\HII}{\ion{H}{2}}
 \newcommand{\cmmb}{cm$^{-2}$}
 \newcommand{\mo}{M$_\odot$}

\shorttitle{\HI\ in NGC~5253}
\shortauthors{Kobulnicky \& Skillman}

\begin{document}

\title{Inflows and Outflows in the Dwarf Starburst Galaxy
NGC~5253: High-Resolution \HI\ Observations }

\author{Henry A. Kobulnicky}
\affil{Department of Physics \& Astronomy \\  1000 E. University \\
University of Wyoming \\ Laramie, WY 82071 
\\ Electronic Mail: chipk@uwyo.edu}

\author{Evan D. Skillman}
\affil{University of Minnesota \\ Department of Astronomy \\ 
115 Church St. SE \\ Minneapolis, MN 55455
\\ Electronic Mail: skillman@astro.umn.edu}


\author{Accepted for Publication in {\it The Astronomical Journal} }


\begin{abstract}

{\it Very Large Array } and Parkes 64 m radiotelescope 21-cm
observations of the starburst dwarf galaxy NGC~5253 reveal a
multi-component non-axisymmetric \HI\ distribution.  The
component associated with the stellar body 
shows evidence for a small amount of rotational
support aligned with the major axis, in agreement with
optically measured kinematics and consistent with the small
galaxian mass.  Approximately 20--30\% of the \HI\ emission
is associated with a second component, an \HI\ ``plume''
extending along the optical minor axis to the southeast.
We consider outflow, inflow, and tidal origins for this feature.
Outflow appears improbable, inflow is a possibility, and
tidal debris is most consistent with the observations.
Thus, kinematics of
the \HI\ that include this feature are not indicative of the
dynamical mass or the local velocity dispersion of the cold
gas.  These \HI\ observations also reveal a filamentary
third component that includes an 800 pc diameter \HI\ shell
or bubble to the west of the nucleus, coinciding with an
H$\alpha$ shell.  The mass of \HI\ in the shell may be as
large as $\sim$$4\times10^6$ \msun.  This large mass,
coupled with the lack of expansion signatures in the neutral
and ionized gas ($v<30$ \kms), suggests that this feature
may be an example of a starburst-blown bubble stalled by
interaction with a massive neutral envelope.  Many other
\HI\ kinematic features closely resemble those seen in
H$\alpha$ emission from the ionized gas, supporting the
interpretation of neutral and ionized gas outflow at
velocities of $\sim$30 \kms.  Comparison between extinction
estimates from the Balmer emission-line decrement and the
\HI\ column densities suggest a gas-to-dust ratio 2--3 times
the Galactic value in this low-metallicity ($Z=1/4~Z_\odot$)
galaxy.
\end{abstract}

\keywords{ 
galaxies: individual (NGC 5253) --- 
galaxies: ISM --- 
galaxies: kinematics \& dynamics ---
galaxies: starburst }

\section{Introduction } 

NGC~5253 is a remarkable starbursting dwarf galaxy with a
prominent minor axis dust lane located in the nearby
Centarus A/M~83
galaxy complex \citep{kara}  .  At visual wavelengths, the
impression of NGC~5253 is dominated by its network of
ionized gas filaments which extend across its minor axis and
reach beyond the stellar distribution \citep{hodge,
graham81, caldwell, marlowe, martin}.  X-ray observations
reveal several small starburst-heated bubbles but no
monolithic superbubble or galactic wind \citep{strickland,
summers}.  It harbors an extremely luminous, compact,
obscured site of nuclear star formation containing several
star clusters with masses from $10^5$ -- $10^6$ $M_\odot$
having ages $<10^7$ yrs \citep{gonzalez, calzetti97, vanzi,
martin-h}.  Much of the nucleus is obscured by high levels
of patchy dust extinction up to $A_V=25$ mag \citep{aitken,
moorwood, alonso}. Radio-wave continuum and recombination
line studies have revealed a dominant luminous ($10^9$
$L_\odot$), compact (r=1 pc), young ($<2.3$ Myr) star
cluster that appears to be gravitationally bound and
potentially destined to evolve into a globular cluster
\citep{beck, turner98, gorjian, mohan}.  The centimeter-wave
spectral energy distribution is consistent with an
optically-thick thermal Bremsstrahlung
source\footnote{\citet{turner00} term this object a
``supernebula'' while \citet{kj} use the name ``ultra-dense
\HII\ regions'' (UDHII) by analogy with the Galactic
ultra-compact \HII\ regions \citet{wood}.}  having densities
exceeding $10^4$ cm$^{-3}$ and emission measures $> 10^8$
cm$^{-6}$ pc powered by $> 10^{52}$ Lyman continuum photons
s$^{-1}$. \citet{caldwell} studied the star formation
history of NGC~5253 and determined that the light was
dominated by a young component with an age between 10$^8$
and 10$^9$ yrs.  Recent studies have concluded that all of
the UV-selected star clusters in NGC~5253 have
estimated ages $<$20 Myr, indicating that either the star
formation episode began very recently or else that the
timescale for dynamical disruption of star clusters is very
short \citep{tremonti, harris}.  The overall gas-phase
metallicity of NGC~5253 is 12+$\log(O/H)\simeq8.16$ (about
1/4 solar), but some of the central regions contain an
excess of nitrogen, consistent with local ``pollution'' from
Wolf-Rayet star winds \citep{wr, k97, ls07}.

When observed at low resolution, the distribution and
kinematics of its neutral interstellar medium suggest that
the \HI\ in NGC~5253 either rotates about the minor axis or
flows radially along the minor axis
\citep{ks}.  \citet{turner97} and \citet{meier} report the
$^{12}CO$ detection of molecular gas clouds having
kinematics consistent with infall along the minor axis dust
lane.  These peculiar gas kinematics may hold clues to the
mechanism that triggered the present burst of star
formation.

In this paper we present 21-cm aperture synthesis
observations from the National Radio Astronomy Observatory's
{\it Very Large Array}\footnote{The National Radio Astronomy
Observatory is operated by Associated Universities,
Inc. (AURA) under cooperative agreement with the National
Science Foundation.} (VLA) and the Parkes 64 m radio
telescope to investigate the conditions in the interstellar
medium that have produced such extreme star formation.
NGC~5253 is an ideal environment to seek the causes of
extreme star formation because the burst is so young and,
presumably, feedback from the energetic star clusters has
not yet had a dramatic impact on the surrounding ISM.  

The distance to NGC~5253 is somewhat controversial and has been measured 
in many studies including 
\citet{saha} (4.1 $\pm$0.5 Mpc),
\citet{gibson} (3.3$\pm0.3$ Mpc), 
\citet{freedman} (3.25$\pm$0.2 Mpc),
\citet{karachentsev} (3.9$\pm$0.5 Mpc), 
\citet{thim} (4.0$\pm$0.3 Mpc),
and \citet{sakai} (3.8 $\pm$0.2 Mpc).
The distances based on Cepheid variables strongly depend on the sample selection
\citep{gibson, thim} and so we prefer the TRGB distance of 
3.8 Mpc \citep{sakai} which implies an
angular scale of 18.4 pc arcsec$^{-1}$.

\section{Observations}

\subsection{VLA}

NGC~5253 was observed in the 21-cm line of neutral hydrogen
using the $VLA$ for approximately 6 hours in the BnA
configuration on 1995 September 25, 3 hours in the CnB
configuration on 1996 January 28, and 2 hours in the DnC
configuration on 1996 June 10.  The 2IF correlator
configuration with Hanning smoothing yielded a total
bandwidth of 1.56 MHz (327 \kms) and 128 12.2 kHz (2.58
\kms) spectral channels.  Intermittent observations of the
compact radio source 1316-336 were used to phase calibrate
the data, and observations of 3C286 were used for flux
calibration.  The data were reduced with NRAO's $AIPS$
package following standard procedures for editing bad time
ranges and performing phase, flux, and bandpass
calibrations.  The radio continuum flux was subtracted in
the UV plane using offline channels to produce a
continuum-free 21-cm line UV dataset.  Data from each
configuration were calibrated separately, and then the UV
data were combined after subtracting continuum.  These
combined data were then used to map the emission from the
1420 MHz transition of neutral hydrogen with greatly
enhanced angular resolution and sensitivity compared to the
45-minute DnC configuration observations presented by
\citet{ks}.

The UV data were mapped to the image plane at two different
resolutions and CLEANed \citep{clark, hogbom} with resulting
clean synthesized beamsizes of 9.0\arcsec$\times$7.6\arcsec\
and 17\arcsec$\times$15\arcsec\ to emphasize both the small
scale structures and diffuse \HI\ features, respectively.
The RMS noise levels in these maps are 1.2 mJy beam$^{-1}$
and 1.7 mJy beam$^{-1}$, respectively, slightly larger than
the theoretical noise of 1.0 mJy.  Because the angular
extent of \HI\ in NGC~5253 ($\sim2$\arcmin; \citet{ks}) is
much smaller than the $VLA$ primary beam size at 21 cm, no
primary beam correction was applied.  The low-resolution
cube was used as a template in which regions having emission
exceeding 2$\sigma$ in at least two adjacent channels were
marked.  The low resolution cube was then used to blank the
higher resolution cube on a channel-by-channel basis outside
the areas of probable emission.  This approach preserves
low-level 1--2 $\sigma$ emission peaks, many of them real,
in the final high-resolution data cube used for subsequent
analysis.

\subsection{Parkes}

NGC~5253 was observed with the Parkes 64 m radio telescope
in the 21-cm line of neutral hydrogen on 1997 February 12
using a total bandwidth of 8 MHz (1600 \kms) and 2048
channels of 0.0039 MHz (0.82 \kms) centered at $v_\odot=250$
\kms.  The primary beam angular diameter is 16\arcmin.  We
interleaved on-band spectra with frequency-switched off-band
spectra centered at $v_\odot= 880$ \kms\ to obtain
emission-free \HI\ profiles.  A 9-pointing (3x3) grid with
14\arcmin\ spacings centered on NGC~5253 was used to search
for extended emission surrounding NGC~5253, but only the
central position showed a strong 21-cm detection. The data
for left and right circular polarizations were reduced by
subtracting the frequency-switched reference spectrum from
the source and fitting a low-order polynomial to remove any
residual baseline.  The system temperature averaged 33 K and
the mean rms noise in the resulting spectra is 0.06 K per
channel.  We observed the Galactic \HII\ region S8 as a flux
calibration standard and scaled the observed temperatures to
match the 76~K reported by \citet{kalberla}.

\section{Analysis of the New \HI\ Observations}

\subsection{Total \HI\ and Line Profiles}

\begin{figure}
\plotone{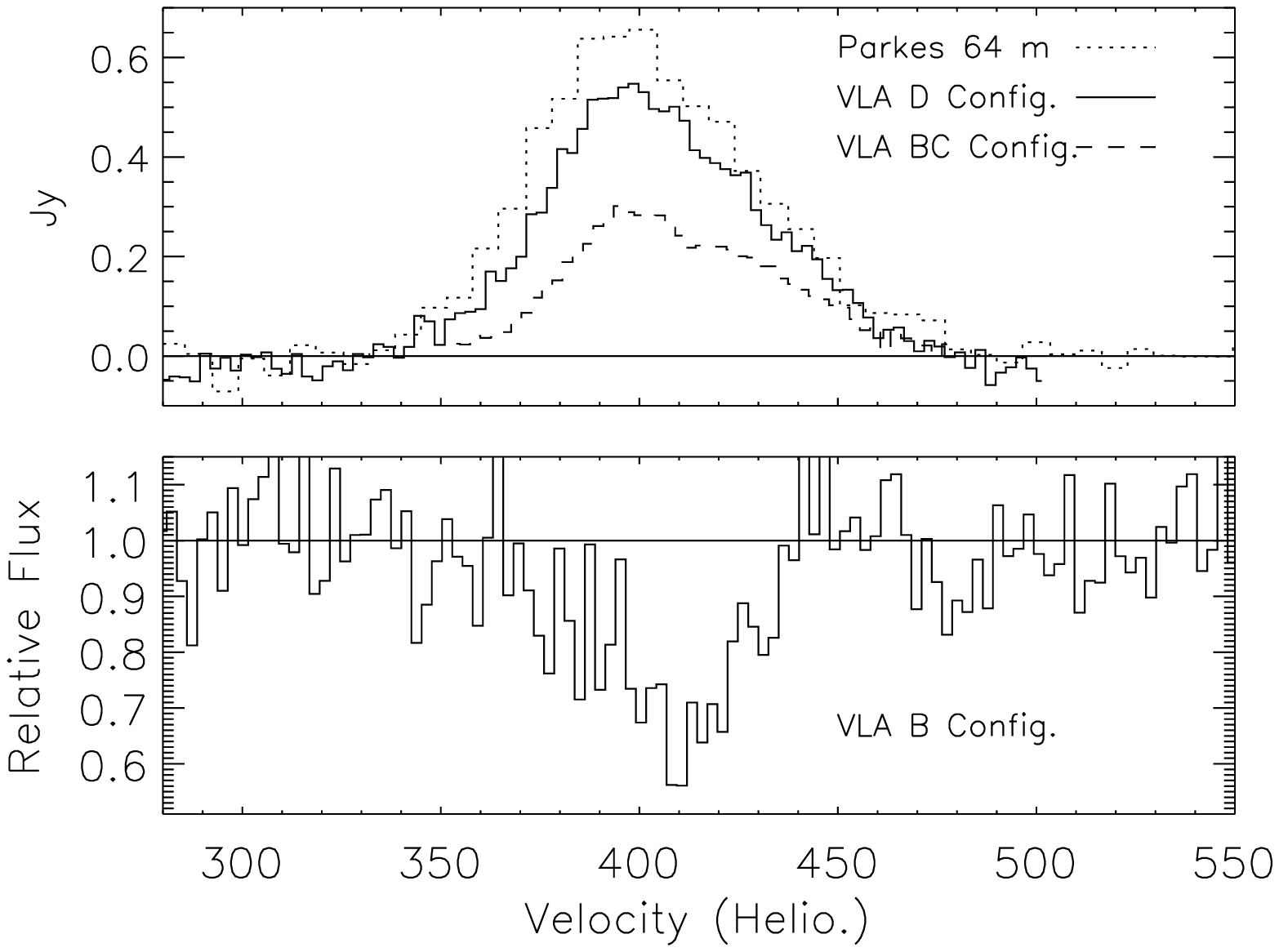}
\caption{21-cm emission line profiles (top) from the Parkes 64 m telescope
(dotted line), the $VLA$ D configuration data (solid line) and the $VLA$ 
B+C configuration data cube (dashed line).   The lower panel shows the 
21-cm absorption line profile from the longest $VLA$ baselines
against the compact thermal 
radio continuum source associated with the super star cluster 
in the nucleus \citep{turner98}.       \label{oned} }
\end{figure}

Figure~\ref{oned} shows the 21-cm emission line profile (top
panel) of NGC~5253 from the Parkes 64 m telescope (dotted
line), the $VLA$ D configuration short-baseline data (solid
line) and the $VLA$ B+C configuration high-resolution data
cube (dashed line) as described above.  The lower panel
shows the 21-cm absorption line profile from the longest
baselines of the $VLA$ B configuration against the compact
thermal radio continuum source associated with the super
star cluster in the nucleus \citep{turner98}.  The
integrated single-dish \HI\ flux is 43.9 Jy \kms, equivalent
to a neutral hydrogen mass of $1.42\times10^8$ \mo\ at a
distance of 3.8 Mpc, assuming,

\begin{equation}
{{M_{HI}} \over {M_\odot}}= 2.36\times10^5 [D(Mpc)]^2 \int S(Jy) \delta v (km~s^{-1}).  
\end{equation}  

The $VLA$ D configuration data recovers 77\% of this total,
implying the presence of some $3\times10^7$ \mo\ of neutral
hydrogen distributed smoothly on scales larger than
$\sim7$\arcmin, the largest structures visible in 2 hours to
the $VLA$ at 21 cm in the D configuration \citep{ulvestad}.  The
high-resolution data cube contains 42\% of the single-dish flux, or about
$6.0\times10^7$ \mo.  Therefore, \HI\ column densities measured 
from this cube will be {\it lower limits} to the true values
which may be larger by a factor of order $\lesssim 2$.  

The 21-cm absorption spectrum in the lower panel of
Figure~\ref{oned} shows a broad absorption line spanning at
least 80 \kms\ with a maximum optical depth of $\tau=0.55$
near $v_\odot=409$ \kms.  The velocity of maximum optical
depth is redshifted $\sim10-20$ \kms\ compared to the 21-cm
emission profile peak at $v_\odot\simeq395$ \kms.  This is
consistent with the central radio supernebula lying behind the
bulk of the cool atomic gas and the latter having a net
radial velocity toward the nucleus, as suggested from the
kinematics of the cold molecular clouds by \citet{meier}.
Adopting \citep{dl},

\begin{equation} 
N_H ={{ \tau~\Delta{v}~T_s} \over {5.2\times10^{-19}}},
\end{equation} 

\noindent where $\tau$ is the 21-cm optical depth, 
$T_s$ is the spin temperature and
$\Delta{v}$ is the velocity full width at half maximum of
the 21-cm absorption line, we can estimate the column
density, $N_H$, of neutral hydrogen in front of the compact
nuclear radio supernebula.  Because the solid angle
probed by the $VLA$
absorption and emission spectra are vastly different, we do not
have an estimate of the spin temperature for this gas, so we
adopt a lower limit of $T_s>50$ K \citep{dl}.  The overall
width of the feature is consistent with an abundance of cold
\HI\ clouds in front of the nucleus over a range of
velocities $v_\odot = 370 - 435 $ \kms.  Therefore, we adopt
a lower limit of $\Delta{v}>10$ \kms.  The resulting \HI\
column densities are $N_H > 5\times 10^{20}$ \cmmb\ at this
location.

\subsection{The \HI\ Distribution at High Resolution}

\begin{figure}
\plotone{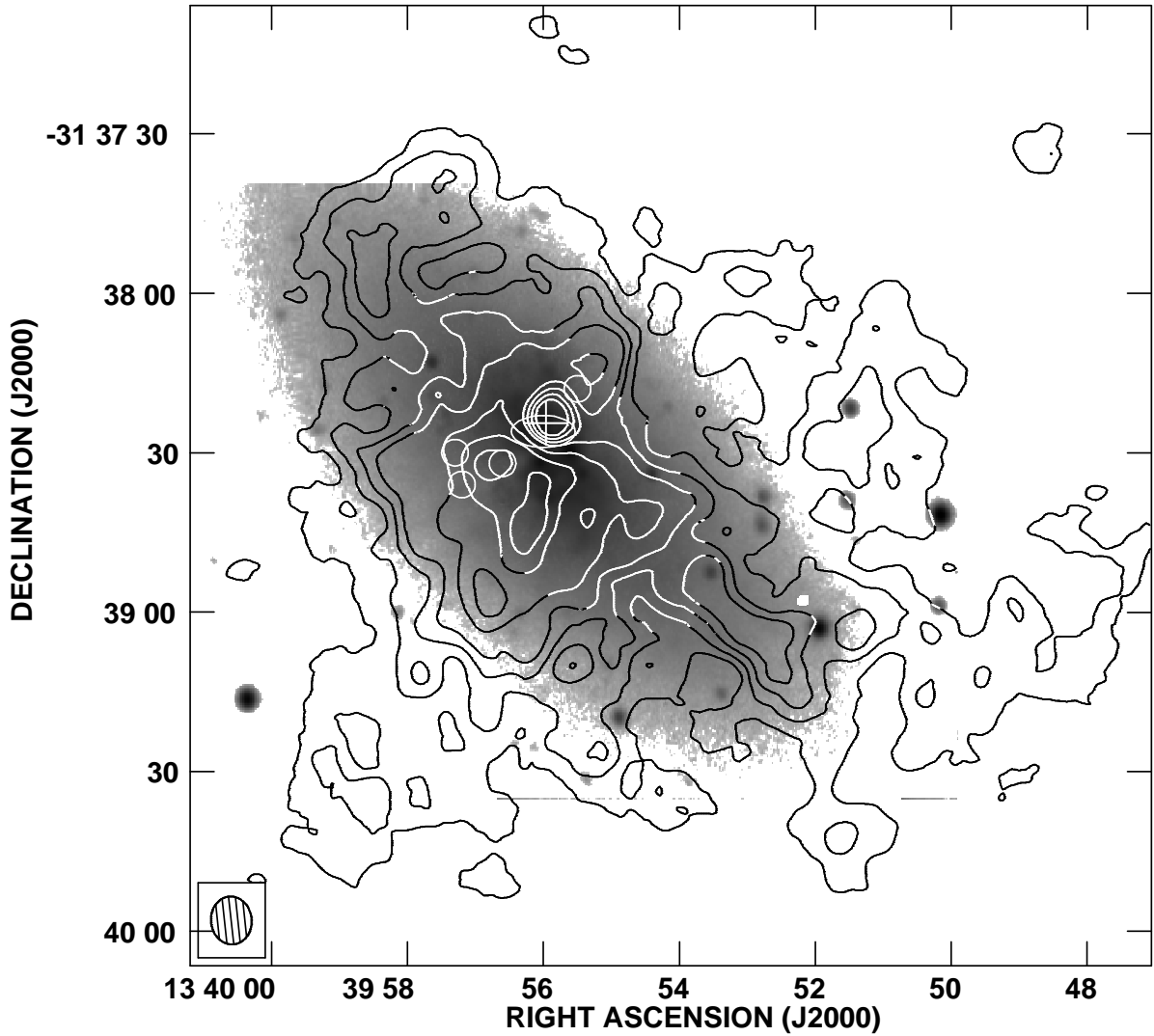}
\caption{NGC~5253 \HI\ total intensity map (contours) with
optical 6450 \AA\ continuum image (greyscale).  Contours levels
show  \HI\ beam-averaged 
column densities of  6, 12, 18, 24, 36, 52, and 64   
$\times10^{20}$ \cmmb.   An 8\arcsec\ diameter (148 pc) 
cross marks the location of the
dominant compact central star cluster and radio ``supernebula''
\citep{turner98}. Note that the contours at the position of this
feature indicate absorption.
Circles mark the positions of CO clouds
reported by \citet{meier}. The inset at lower left shows
the size of the 9.0\arcsec$\times$7.6\arcsec\ synthesized $VLA$ beam.
   \label{mom0} }
\end{figure}

Figure~\ref{mom0} shows the integrated \HI\ column density
(contours) overlaid on a 6450 \AA\ optical continuum image
of NGC~5253 (greyscale).  The inset at lower left depicts
the 9.0\arcsec$\times$7.6\arcsec\ synthesized beam.
Contours show 21-cm line fluxes of 0.04, 0.08, 0.12, 0.16,
0.24, 0.36, and 0.40 Jy \kms\ corresponding to beam-averaged
\HI\ column densities of 6, 12, 18, 24, 36, 52, and 64   
$\times10^{20}$ \cmmb.  A cross marks the location of the
dominant embedded star cluster and thermal radio supernebula
\citep{turner98} where
\HI\ appears in absorption.   The \HI\ column density peaks near
6.4$\times10^{21}$ \cmmb\ at $\sim$15\arcsec\ (270 pc)
south-southeast of radio supernebula and near the dust lane
which coincides with the position of several molecular
clouds \citep{meier}.

As noted by \citet{ks}, the \HI\ distribution in NGC~5253
does not closely follow the optical morphology.  Rather, the
\HI\ is extended along the minor axis to the southeast and
northwest.  Note that at this higher resolution, it is clear
that the extended gas is not symmetrically placed along the
minor axis.  The extension to the southeast is a much more
prominent feature, and we designate this as the \HI\
``plume''.  These new data also reveal arc-like extensions
of \HI\ to the west of the nucleus, suggestive of a partial
shell or bubble of neutral hydrogen.  There are also
extensions of \HI\ to the NE and SW that generally follow
the optical major axis.

\begin{figure}
\plotone{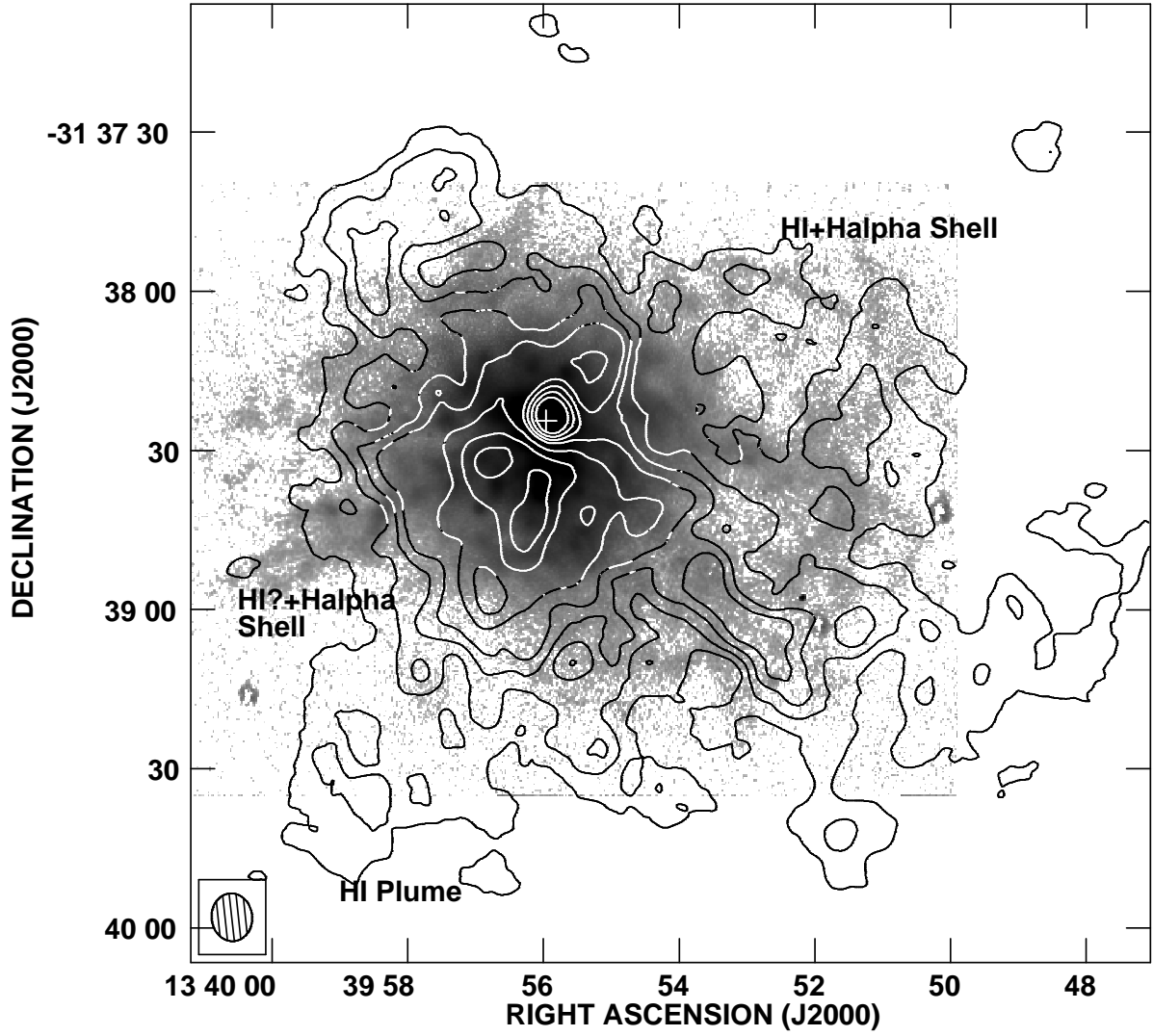}
\caption{\HI\ column density of NGC~5253 (contours) 
as in Figure~\ref{mom0} overlaid on an $H\alpha$
image (greyscale).  Note the correspondence between the H$\alpha$ shell
and \HI\ arcs to the west of the nucleus.
  \label{mom0HA} }
\end{figure}

Figure~\ref{mom0HA} depicts the integrated \HI\ column
density (contours) overlaid on an $H\alpha$ image
(greyscale).  Here we note that the morphology of the \HI\
shell to the west corresponds closely to the morphology of
the $H\alpha$ shell at position angle of about $-$70\degr\
as seen from the central radio supernebula.  The morphology of
the \HI\ to the east appears to mimic the H$\alpha$ emission
in the inner parts of the galaxy where the \HI\ peak column
densities coincide with the H$\alpha$ peaks.  Also in the
east, similar to in the west, there is an H$\alpha$
extension that appears to be outlined in \HI\ (at a position
angle of about 100\degr ). However, further from the center,
this correlation breaks down.  Specifically, the bright \HI\
plume, at a position angle of about 140\degr\ does not have
a corresponding strong H$\alpha$ emission feature.

\begin{figure}
\plotone{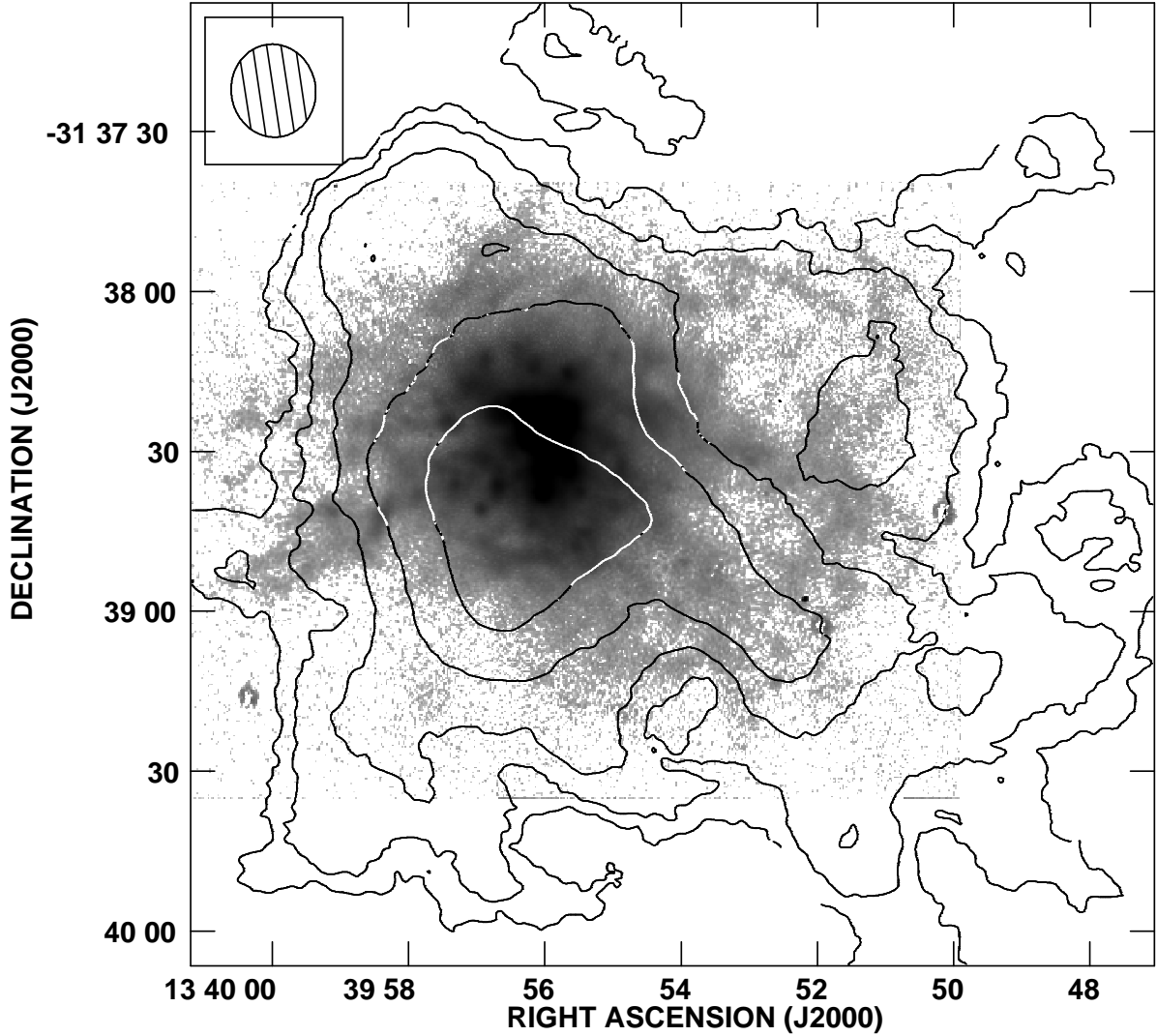}
\caption{\HI\ column density of NGC~5253 (contours and greyscale) 
at a lower resolution than in Figure~\ref{mom0} in order to highlight
the lower column density features.  Contours correspond to beam-averaged
\HI\ column densities of 2, 4, 8, 16, and 32 $\times10^{20}$ cm$^{-2}$.
 Note the \HI\ extensions to the east and west
that coincide with $H\alpha$ filaments.  
\HI\ extensions
along the major axis to the southwest also have $H\alpha$ counterparts.
  \label{MOM0V2} }
\end{figure}

Figure~\ref{MOM0V2} depicts the integrated \HI\ column
density (contours and greyscale) at a lower resolution of
16\arcsec\ useful for identifying lower column density
features.  A few additional features appear in this figure.
The \HI\ extension to the west now appears more clearly and
shows the same open shell geometry as the H$\alpha$.  At
this lower resolution, it is now possible to see that there
is \HI\ at a large radius from the center but on the same
position angle as the center of the western open shell.
This looks reminiscent of simulations of bubbles that
achieve break-out \citep{maclow}.  Also, at lower
resolution, the eastern extension looks more like the same
open shell morphology and also follows the H$\alpha$.

In summary, the integrated \HI\ maps reveal three neutral
gas components that will become more clear in the velocity
channel maps below: 
1) extensions to the NE and SW that follow the optical major axis,
2) an \HI\ plume to the SE that extends well beyond the optical extent of the galaxy, and
3) open shells in the west and east, and 

\subsection{The \HI\ Kinematics}

Figure~\ref{greypanels} shows 21-cm maps of NGC~5253 in
every 2.58 \kms\ velocity channel containing emission from
$v_\odot=459$ to $v_\odot=358$ \kms.  The greyscale range
shows beam-averaged column densities of 1.0$\times10^{20}$
\cmmb\ (white), equivalent to 2$\times$ the rms noise, to
5.0$\times10^{20}$ \cmmb\ (black).  The cross marks the
location of the central super star cluster and radio
supernebula.  Ellipses indicate the positions of the
molecular clouds A through E (from left to right, in
decreasing Right Ascension) identified by \citet{meier}.  In
Figure~\ref{greypanels}, the southeast \HI\ plume is the
dominant feature at higher velocities (450 - 420 \kms ),
while the \HI\ aligned with the stellar distribution is
found primarily at velocities between 420 and 380 \kms .

The kinematics of individual \HI\ features relative to the
ionized gas can be better seen in Figure~\ref{panels}
which shows \HI\ column density channel maps (contours)
overlaid on an $H\alpha$ image (greyscale).  Every second
2.58 \kms\ channel is plotted.  The contours show fluxes of
2.4, 3.6, 6.0, and 8.4 mJy beam$^{-1}$, equivalent to
beam-averaged \HI\ column densities of 1.0, 1.5, 2.5, and
3.5 $\times10^{20}$ \cmmb.  Again, it can be seen that the
highest velocity gas at $v_\odot = 420-455$ \kms\ (upper 2
rows of panels) is located to the southeast of the nucleus.
Gas at intermediate velocities $v_\odot=400-420$ \kms\ near
the systemic velocity of $v_\odot\simeq400$ \kms\ has a
circularly symmetric morphology similar to that of the
$H\alpha$.  At these velocities there are also lower column
density \HI\ extensions along the major axis of the galaxy
to the northeast, southwest and also in the direction of the
H$\alpha$ shell to the west.  At the lowest velocities,
$v_\odot < 400$ \kms, the \HI\ morphology becomes elongated
along the galaxy's major axis.

Figures~\ref{mom0HA} and \ref{panels} also show
\HI\ extensions to the northeast and southwest along the galaxy's
major axis.  The \HI\ protrusion to the northeast appears at
velocities $v_\odot=415-380$ \kms\ and does not
unambiguously form a coherent shell or bubble.  The \HI\
extension to the southwest appears over a similar velocity
range with more gas closer to the systemic velocity.  The
spatial and kinematic symmetry of these extensions along the
major axis is consistent with low amplitude rotation,
in the canonical manner, about the galaxy's minor axis, but
the pattern is, by no means, obvious.

Figure~\ref{spider} shows a 21-cm intensity-weighted
velocity map ( 1st moment; contours) overlaid on a stellar
continuum 6450 \AA\ image (greyscale).  Isovelocity contours
are labeled in \kms.  This figure shows a steep kinematic
velocity gradient along the southeast \HI\ plume.\footnote{The
gradient is nearly parallel to the projected direction toward
the much larger group member M~83, and \citet{vdbergh}
suggested a recent interaction between the two.  However,
the line-of-sight separation exceeds 1~Mpc given the current
distance determinations of 3.8$\pm$0.2 Mpc for NGC~5253 and
5.16$\pm0.41$ Mpc for M~83 \citep{kara}, effectively
precluding any recent interaction.}  The contours show the
largest radial velocities of $v_\odot\simeq450$ \kms\ at a
distance 1\arcmin\ to the south-southeast of the nucleus.
Radial velocities approach the systemic velocity of
$v_\odot\simeq395$ \kms\ near the nucleus.  Along the major
axis to the northeast and southwest of the nucleus, the
velocity field is disordered and shows a small velocity
spread, ranging between $v_\odot\simeq385$ in the northeast
and $v_\odot\simeq405$ \kms\ in the southwest.

\citet{ks} concluded, on the basis of low-resolution (45\arcsec)
\HI\ maps, that the neutral atomic medium 
in NGC~5253 rotates about the {\it major} axis, as in polar
ring galaxies.  However, these new high-resolution \HI\ data
reveal a strong asymmetry in the \HI\ distribution and
velocity field, making the rotation scenario improbable.
These new data are more consistent with an either an inflow or
outflow of neutral gas along the minor axis from the south
and east of the nucleus.  Inflow, was considered by
\citet{ks} and propounded by \citet{turner97} and
\citet{meier} on the basis of  $^{12}CO$ observations.
  
Figure~\ref{spider2} shows the 21-cm intensity-weighted
velocity (1st moment) map as in Figure~\ref{spider}
(contours) overlaid on an \HI\ total intensity map (0th
moment; greyscale).  The scale bar shows \HI\ surface
brightness in units of Jy beam$^{-1}$ m s$^{-1}$.  A peak
flux of 400 Jy beam$^{-1}$ m s$^{-1}$ corresponds to \HI\
column densities of 6.4$\times10^{21}$ \cmmb.  The peak \HI\
column densities occur where the velocity gradient is the
steepest.  The
majority of the \HI\ by mass is located to the
south-southeast and in the circumnuclear region.
\HI\ extensions to the west, southwest, and northeast have 
comparatively low total column densities and low
line-of-sight velocity dispersions, consistent with gas
associated with the main body of the galaxy.

Position-velocity diagrams allow another perspective on the
observations.  Figure~\ref{lvmajor} shows a
position-velocity diagram of the \HI\ emission taken along
the optical major axis at position angle 50\degr.  Although
the diagram is quite complex, two main features stand out.
First, there is a large velocity range in the central parts.
Here one can appreciate the asymmetry in the field.
Redshifted \HI\ gas extends out to 50 \kms\ from the
systemic velocity, while blueshifted gas extends out only 30
\kms.  The second notable component is the gas at roughly
the systemic velocity showing a slow drift in central
velocity across the spatial distribution with an amplitude
of roughly 20 \kms, consistent with that 
expected for such a low-mass galaxy.

Figure~\ref{lvminor} shows a position-velocity diagram of
the \HI\ emission taken along the optical minor axis at
position angle 140\degr.  Here we again see a large range in
velocities over a relatively small spatial extent.  It is
interesting that in this projection, the gas appears
continuous in velocity from the extreme of the redshifted
gas to the extreme of the blueshifted gas.

\section{Interpretation of Major \HI\ Components in NGC~5253}

Here we will try to draw a coherent picture of the three
major components of the \HI\ distribution: 
the \HI\ associated with the stellar distribution,
the southeast \HI\ plume, 
and the open shells to the west and east.

\subsection{\HI\ Along the Major Axis}

It is interesting to compare these new \HI\ observations of
NGC~5253 to those of NGC~625 \citep{cannon04}, another dwarf 
starburst galaxy at a similar distance.  \citet{cote} 
showed that NGC~625 also has
the dominant \HI\ velocity gradient along the {\it minor} axis.
\citet{cannon04} showed that this large velocity gradient
was due to outflowing \HI\ gas from the central starburst.
This outflowing gas is also detected in the warm and hot
phases of the ISM \citep{cannon05}. 

This similarity with NGC~625 motivates a second look at
the \HI\ component in NGC~5253 that is aligned with the optical
galaxy.  In NGC~625, the signature of rotation along the
major axis was swamped by the presence of the strong
\HI\ outflow.  However, the position velocity diagrams
revealed the low amplitude rotation curve present in 
the \HI.  It is probable that we are seeing
the same effect in NGC~5253. Although the \HI\ in NGC~5253
is quite disturbed, the overall trend for higher velocities
in the southwest and lower velocities in the northeast
suggest a pattern of rotation.  This is also seen in the
stellar velocities measured by \citet{caldwell} (their
Figure 10) and the H$\alpha$ major-axis velocity map of
\citet{martin} (her Figure 9).  It is not straightforward to
calculate the expected amplitude of the rotation curve for
NGC~5253;  comparing to the luminosity requires a significant
correction for the small M/L ratio expected for a starburst galaxy.
Nonetheless, after applying a small ($\le$ 20\%) correction for
inclination, the amplitude of $\sim$ 10 \kms\ is not out of
line with what could be expected for NGC~5253.  
Note that the \HI\ study of the similar dwarf starburst galaxy
NGC 1569 by \citet{stil} also found only small evidence
of normal rotation in the
inner parts of NGC~1569, but that the \HI\ along the major
axis in the outer parts confirmed this rotation signature.

\subsection{The \HI\ Plume}

These new high resolution \HI\ observations preclude 
rotation along the minor axis as a viable
interpretation for the southeast plume.  Either inflow, outflow,
or some kind of tidal interaction remnant feature
seem more likely. Based on CO observations,
\citet{turner97} favored infall of metal poor gas.
The radial velocities of the CO clouds $\sim$200 pc east of  the
nucleus are slightly higher than the systemic velocity, consistent with
inflow or possibly rotation about the major axis within the 
innder few hundred pc.  The \HI\ velocities at this location could
also be consistent with major-axis rotation within the nuclear
regions.  Many galaxies are known to have counter-rotating or
decoupled dynamical components in the inner regions, and close
grvitational interactions are often invoked as the causes of such
kinematics.

That
an accretion event may have triggered the current starburst
has been suggested several times \citep{vdbergh, graham81,
caldwell}. \citet{turner97} identify three points in favor
of the infall hypothesis.  First, the weak CO emission
associated with a strong starburst favors low metallicity
gas.  Second, the non-central position of the CO suggests a
transient feature.  Third, the redshift of the CO relative
to the galaxian systemic velocity could be indicative of
inflow.
\citet{meier} further note that since the dust lane that is
associated with the CO and \HI\ emission in the southeast plume
appears to be in front of the galaxy, the geometry
(redshift) favors infall.  

A close look at the relative locations of the CO clouds,
dust lane, and \HI\ peaks along the southeast minor axis
reveals, in fact, two different gaseous components in
this direction.  These two components are resolved both 
spatially and kinematically by the
new \HI\ observations, and a comparison with the 
extinction maps of \citet{calzetti97} shows that these are
intrinsically different components --- not two maxima of
the same feature.

Figure~\ref{AV} shows a greyscale
representation of the dust reddening ($A_V$) computed from
the $H\alpha/H\beta$ ratio maps of \citet{calzetti97}
assuming $A_V=3.1 E(B-V)$.  The inferred extinctions range
from $A_V(H\alpha/H\beta)$ near zero to
$A_V(H\alpha/H\beta)\simeq2$.  The cross and ellipses show
the locations of the central thermal radio supernebula and the
CO clouds, as in previous figures. Contours in
Figure~\ref{AV} show the extinction derived from the neutral
hydrogen column densities assuming a Galactic ratio of gas
to dust \citep{bohlin},

\begin{equation}
\left< {{(N_{HI} +2N_{H_2})} \over {E(B-V)}} \right> = 5.8\times10^{21} 
atoms~cm^{-2}~mag^{-1}
\end{equation}  

\noindent For a standard Galactic ratio of total to selection extinction, 
$R\equiv A_V/E(B-V) =3.1$, this is equivalent to,

\begin{equation}
3.1 \left< {{(N_{HI} +2N_{H_2})} \over  {5.8\times10^{21} 
atoms~cm^{-2}~mag^{-1}}} \right> = A_V
\label{dustgas}
\end{equation}  

Here, we have ignored, for the moment, any contribution from
molecular gas, and we remind the reader that the \HI\ column
densities inferred from the high-resolution interferometer
maps may be underestimates of the true column densities by
up to a factor of two, owing to the insensitivity of the
$VLA$ to emission on large angular scales.  Hence, the
estimated \HI\ column densities and extinctions are lower
limits.  The contours in Figure~\ref{AV} correspond to
$A_V(HI)=$4, 3.5, 3, 2.5, 2, 1.5, 1 and 0.5 mag.

Figure~\ref{AV} shows that the extended region of high
extinction indicated by the Balmer decrement coincides with
a local maximum in the neutral hydrogen column density and
the locus of the positions of the CO detections.  The ratio
$A_V(HI)/A_V(H\alpha/H\beta)$ in this region $\sim
3.5/2=1.75$, consistent with the expectation that some of
the \HI\ lies behind the region of ionized gas, and with the
expectation that some of the diffuse \HI\ may be unseen in
the interferometer maps.  More precise estimates are not
possible because of the large disparity in beamsize between
the $VLA$ \HI\ data (8\arcsec) and the \citet{calzetti97}
$HST$ maps (0.3\arcsec). Nevertheless, the correlation
between both extinction measures at this location is
consistent with the majority of the \HI\ (and molecular
gas?)  lying in the foreground of the ionized medium.  The
high \HI\ column density in this feature probably continues
into the region surrounding the central radio supernebula and
super star cluster, but measurements are unreliable at this
location owing to the strong \HI\ absoprtion.

By contrast, the maximum \HI\ column density, corresponding
to $A_V(HI)=4$ mag, is located in the \HI\ plume $\sim$15\arcsec\ to the southwest
of the region with the largest $A_V(H\alpha/H\beta)$.  At
this location there is no pronounced peak in
$A_V(H\alpha/H\beta)$ and there are no CO detections.
Figure~\ref{greypanels} shows that the gas responsible for
this primary peak in N(\HI) occurs at velocities
$v_\odot>410$ \kms\ and is kinematically and spatially
distinct from the neutral gas at the location of the
molecular clouds.  {\it Thus, what we are calling the \HI\ plume
appears to have no physical relation to the region of high
extinction in the dust lane}, with the possible exception of
a tenuous connection at velocities $v_\odot<225$ \kms.

The very high ratio of \HI\ column density to extinction in the
plume implies that either the plume gas is located on the
far side of NGC~5253, or that the gas is extremely
metal-poor and has a very large gas-to-dust ratio.  The
metal-poor ISM in dwarf galaxies does exhibit gas-to-dust ratios
2--5 times higher than in the Milky Way \citep{draine}, so
equation~\ref{dustgas} could require a correction factor of
this order.  However, the mean ratio
$A_V(HI)/A_V(H\alpha/H\beta)=3.75/0.5=7.5$ in the vicinity
of the plume compared to $A_V(HI)/A_V(H\alpha/H\beta)=3.25$
near the dust lane suggests that the plume and the dust lane 
are distinct entities.   Even if we allow that, statistically,
half of the neutral medium may lie behind the star-forming
ionized gas, the inferred extinctions are factors of 2--3
larger from the \HI\ column density than from the Balmer
decrement.  The real ratio may be even larger because, as
described above, the \HI\ column density estimates should be
regarded as lower limits and a molecular component is not
included.  

The true nature of this low-extinction, high-column-density,
redshifted \HI\ plume is difficult to pin down.  One
possible explanation is that the \HI\ plume may be
outflowing gas such as commonly observed in dwarf starbursts
like NGC~1705 \citep{meurer98} and NGC~625 \citep{cannon04}.
The main difference between NGC~5253 and NGC~625 is the
asymmetry of the \HI\ distribution seen here.  The
symmetrical \HI\ distribution in NGC~625 was taken as strong
evidence against infall of gas, but the \HI\ plume in
NGC~5253 does not show a similar
symmetry. Figure~\ref{lvminor} shows that there is
continuity from the strongly redshifted gas ($\sim$ +50 km
s$^{-1}$) through to blueshifted gas ($\sim$ $-$30 km
s$^{-1}$), which would be the signature of somewhat
asymmetric outflow.  The ``spur'' of outflowing gas in
NGC~1705 is one-sided, and \citet{meurer98} note that this
is to be expected when the source of the outflow is
displaced relative to the midplane \citep{maclow89}.
However, the plume in NGC~5253 contains 20--30\% of the \HI\
seen in the interferometer maps, depending on the spatial
and velocity boundaries adopted.  Such a fractionally
massive outflow of neutral gas over a fairly small solid
angle without any ionized gas counterpart makes the outflow
interpretation less plausible in this galaxy.
 
Radial inflow, suggested by \citet{meier} on the basis of
the kinematics of CO clouds within the minor axis dust lane,
is another possibility.  The geometry of the isovelocity
contours in the redshifted \HI\ plume can help distinguish
whether the inflow could be on the near or far side of NGC~5253.
The most redshifted contours (455 \kms) lie farthest from
the galaxy and are $\sim$50 \kms\ larger than the systemic
velocity of 400 \kms.  An infalling stream is expected to
accelerate as it approaches the galaxy.  Thus, if the plume
were infalling gas on the {\it near} side, the isovelocity
contours closer to the galaxy ought to exceed 455 \kms.
Instead, we observe that they approach the systemic
velocity.  Inflow from the {\it far} side of NGC~5253 might,
however, produce the observed signature.  Such a background
geometry would simultaneously explain the lack of extinction
seen toward the high column density peaks in the plume.

Finally, and perhaps most plausibly, the \HI\ plume may be a
kinematically distinct remnant from a recent gravitational
interaction with another member of the Centarus A subgroup
of the Centarus A/M~83 complex.  It is difficult to prove or
rule out this possibility.  In this scenario, the \HI\ asymmetry is easily
understood as a chance projection against the main body of
NGC~5253, and there is no need for the tidal tail to be
symmetrically placed in either space or velocity.  
The galaxy density within the Centarus A subgroup
is high, and interactions are to be expected.  
Many of these systems are dynamically disturbed systems,
suggestive of a history fraught with close encounters.
A perturbing, but not destructive, close encounter 
would simultaneously provide an explanation for the 
\HI\ plume as a tidal feature and provide a cause for the
recent burst of star formation seen throughout the galaxy,
most exceptionally in the nucleus. 

 Karachentsev et al.\ (2007) have provided a three dimensional
picture of the Cen A/M83 complex, and NGC~5253 is located in the
periphery.  Using radial velocities and accurately measured distances,
Karachentsev et al.\ (2007) have identified NGC~5253 with the Cen A
sub-group based on a positive tidal index.  This is because, even though
NGC~5253 is within 2 degrees of M~83 in projection on the sky, it is
1.4 Mpc closer than M83 and nearly identical in distance to Cen A.
Note, however, that NGC~5253 is separately by roughly 12 degrees in
projection from Cen A, corresponding to roughly 800 kpc. Thus NGC~5253
is currently far from the strong gravitational influence of either
of the largest galaxies in the complex.   Despite being a member of
a very populous complex, NGC~5253 appears to be relatively isolated.
In fact, the next nearest known galaxy, ESO 383-087, appears to be
over 400 kpc away.  It is not clear which member of the
group might have interacted recently with NGC~5253.
 
In some ways, the \HI\ plume in NGC~5253 is reminiscent of
some of the extended filaments of gas surrounding IC~10
\citep{ss89, wm98} or the giant \HI\ cloud in Virgo, \HI\ 1225+0146
\citep{mcmahon}.  In IC~10, \HI\ ``plumes'' and 
``streamers'' have velocity gradients that show little relation
to the overall rotation of the galaxy.  The true nature
of the extended \HI\ features in IC~10 is also not yet understood,
but interaction with these \HI\ features is generally thought 
to have provoked the present starburst.  \citet{taylor}
finds that starbursting dwarf galaxies like NGC~5253
are found preferentially to have close companions, so the interaction
hypothesis is a reasonable one.  

One very important point is that the large range of
velocities observed in the \HI\ in NGC~5253 is associated
with the southeast plume.  It is common practice in
the literature to apply the FWHM of the gas velocity
distribution as indicative of internal dynamics to estimate
dynamical mass \citep{harris} or a stellar crossing time
\citep{calzetti97}.  Regardless of whether the large-scale
velocity gradient seen in the \HI\ plume is due to infall,
outflow, or tidal debris, the resulting large velocity dispersion should not
be used as an internal dynamical indicator.

\subsection{The \HI\ Shells and Bubbles}

The main question concerning the open shells on the west and
the east is whether those shells represent events centered
at the currently indicated positions or if they are also the
result of centrally energized outflows that give rise to
shell geometries as they emerge from the galaxy.  
The centers of the western and eastern open shells
lie well outside of the main stellar distribution and far
from the current, centrally concentrated star formation.
Comparison with simulations of outflows \citep{maclow} and
real outflows observed in H$\alpha$ and X-rays
\citep{strickland04} show that open-shell morphologies can
be created where the centers of the shells are far from the
source of the energy.  Is it possible that this is what we
are seeing in NGC~5253?

How likely is it that star formation events capable of
creating large \HI\ shells could have occurred on the
periphery of the present stellar distribution?  Our view of
the spatially resolved star formation history of NGC~5253
has evolved with time \citep{vdbergh, caldwell, calzetti97,
tremonti, harris}. The most recent work by \citet{harris}
emphasizes the very young ages ($\le$ 20 Myr) for all of the
clusters that they studied
\citep[in agreement with the work by][]{tremonti}.  However,
\citet{harris} required detection in the UV for inclusion,
so their sample should be strongly biased
toward the youngest clusters (only the brightest of the
older clusters would be detected - see their Figure 15).
The earlier study by \citet{calzetti97} derived ages of
order 50 Myr for some of the clusters, the colors of the
core required starburst activity within the last 100 Myr,
and \citet{caldwell} deduced a range in cluster ages with an
upper limit of 10$^9$ years.  From the above, it is not
clear whether strong star formation events in the periphery
of its stellar distribution in the recent past are
consistent with the star formation history of NGC~5253.  A
star formation history derived from the field stars, while
challenging owing to the distance and the presence of
differential extinction, would be a valuable study.

Figure~\ref{mom0HA} depicts the integrated \HI\ column
density (contours) overlaid on an $H\alpha$ image
(greyscale).  The morphology of the \HI\ shell to the west
corresponds closely to the morphology of the $H\alpha$ shell
at position angle of about $-$70\degr\ as seen from the
central radio supernebula.  The shell has an angular
diameter of $\sim$45\arcsec\ or about 830 pc.  This
H$\alpha$ feature lies along echelle slit \#2 of
\citet{martin} who noted the absence of kinematic splitting
in $H\alpha$ that would normally demarcate an expanding
ionized shell.  \citet{martin} characterizes this as a
quiescent shell.  The mass of neutral hydrogen in this \HI\
feature is $\sim4.8\times10^6$ \msun\ as measured by summing
emission in the high-resolution data cube from this feature.
Such a significant mass of neutral gas could be responsible
for stalling the shell's expansion and preventing
``blowout''.  \citet{calzetti04} show emission-line ratio
maps for the ionized gas in this vicinity, and they
concluded on the basis of high [S~II]/H$\alpha$ ratios that
the numerous shells and radial filaments to the south and
west are shock ionized, consistent with starburst powered
shells.  However, a softer radiation field, characteristic
of the diffuse ionized gas at large distances from star
forming regions, could also explain these line ratios
\citep{martin97,hunter}.

In Figure~\ref{panels} it can be seen that the
\HI\ coincident with the western $H\alpha$ shell
spans a range of velocities $v_\odot=380 - 410$ \kms.
However, the fragmentary morphology of the neutral gas at
these velocities does not provide clear evidence of a single
coherent structure.  The velocity range (30 \kms) of the
\HI\ near this H$\alpha$ shell is comparable to that found
in the ionized gas, consistent with slow or stalled
expansion.  The lack of X-ray emission inside this bubble
\citep{summers} is also consistent with it being an older,
cooler, quiescent remnant from bygone era of star formation.
If we take the mass of the \HI\ bubble to be $4.8\times10^6$
\mo\ and the radial expansion speed to be 20 \kms\, the
kinetic energy contained in this structure is
$1.7\times10^{52}$ erg.  The implied expansion timescale
for this $\sim$1 kpc diameter structure is
$<$40 Myr, making it unlikely that the current
3 Myr old burst produced this bubble.  It is probably
a product of the older episodes of star formation
from the previous 50 Myr.   
\citet{marlowe} estimate the
mechanical energy injection rate for NGC~5253 to be
$3.6\times10^{40}$ erg s$^{-1}$, or about $5\times10^{54}$
erg over the last 5 Myr.  Even though star formation and its
effects are distributed throughout NGC~5253, this energy
injection rate is more than sufficient to power the observed
\HI\ shell, presuming that such an injection rate has persisted for
at least 50 Myr---the ages of the clusters studied by  \citet{calzetti97}.
     
\section{Comparison of the \HI\ with Other Gas Phases} 

\subsection{Comparison with the Molecular Gas}

Figure~\ref{greypanels} shows the positions of CO detections
identified by \citet{meier} as molecular clouds A through E
(from left to right, in decreasing Right Ascension).
Molecular cloud D is nearly coincident with the central
radio supernebula (marked by a cross).  \HI\ appears
in absorption at this location, so it is not possible to
identify molecular cloud D with a particular 21-cm feature.
Molecular clouds B ($v_{LSR}=421$ \kms) and C ($v_{LSR}=419$
\kms) have probable \HI\ counterparts seen in
Figure~\ref{greypanels} between $v_{\odot}=430$ \kms\ and
$v_{\odot}=405$ \kms.\footnote{In the direction of NGC~5253,
the radial velocity in the heliocentric velocity frame is
nearly identical (within 1 \kms) to the Local Standard of
Rest frame, so we do not explicitly transform our 21-cm data
into the LSR frame used by \citet{meier}.}  Molecular cloud
A ($v_{LSR}=363$ \kms) has no apparent 21-cm counterpart in
a similar velocity range. However, \HI\ emission is present
at that location over the velocity range $v_{\odot}=418-395$
\kms.  Molecular cloud E ($v_{LSR}=424$ \kms) has no
corresponding \HI\ counterpart in a similar velocity range.
Detectable 21-cm emission is present at velocities
$v_{\odot}=410-380$ \kms.  \citet{meier} note that clouds D
and E are tentative detections at the 3--5 $\sigma$ level of
significance, and they may not be real.  The lack of an \HI\
counterpart at the velocity of cloud E is consistent with it
being noise in the CO data.
    
The peak \HI\ column density of 6.4$\times10^{21}$ \cmmb\ measured here
can be used together with the peak $H_2$ column density measurement of
\citet{meier} to make a better estimate the extinction, $A_V(HI+H_2)$,
toward the dust lane, as described above.  
The column density of molecular hydrogen can be estimated 
from the integrated $^{12}CO$ emission, $I_{CO}$, assuming that 
$I_{CO(2-1)}/I_{CO(1-0)}\simeq1$ \citep{meier01}, and adopting
\citep{bloemen, scoville},

\begin{equation}
N_{H_2} (cm^{-2}) \simeq 3\times10^{20} I_{CO} (K~km~s^{-1}).
\end{equation}  

\noindent In terms of the CO flux, $F_{CO}$,

\begin{equation}
N_{H_2} (cm^{-2}) \simeq 3\times10^{20} G~(K~Jy^{-1}) ~F_{CO} (Jy~km~s^{-1}),
\end{equation}  

\noindent where $G$ is the effective gain of the aperture
synthesis beam, approximately 0.46 K Jy$^{-1}$ for the
beamsize and frequency of the \citet{meier} {\it Owens
Valley Radio Observatory} ($OVRO$) data at 230 GHz. The peak
CO fluxes for the four molecular clouds to the east of the
nucleus measured by
\citet{meier} range from 6.9 -- 16.9 Jy \kms, corresponding to
molecular hydrogen column densities $N_{H_2}=0.95$ --
$2.33\times10^{21}$ \cmmb.  This may be a lower limit to the
true molecular column density if the CO-to-$H_2$ conversion
factor in the low-metallicity gas of NGC~5253 is larger than
in the Milky Way, i.e., $X_{NGC~5253}/X_{MW}>1$ 
\citep[e.g.,][]{maloney, verter, israel}.  
The maximum implied visual extinctions, based on peak
$\sim$8\arcsec\ beam-averaged
\HI\ and $H_2$ column densities are $A_V=4.4$ -- $5.9$.
However, as noted by \citet{calzetti97} and evidenced by
Figure~\ref{AV}, the extinction derived from optical
emission line ratio reddening maps is highly variable over
small scales.  These values range from $A_V\simeq0.5$ mag to
$A_V=2$ mag toward most of the star clusters.  $A_V$ may be $>9$ mag
toward the central embedded cluster, consistent with
estimates of $8$ mag $<A_V<25$ mag from the 9.7 $\mu$m silicate feature
optical depth \citep{aitken}. Because the \HI\ and CO
interferometer measurements provide only beam-averaged
extinction estimates over $\sim100$ pc scales, localized
regions may exhibit larger extinctions than the 4.4 -- 5.9
mag obtained from this approach.

\subsection{Comparison with the Hot Ionized Gas}

Figure~\ref{color} is a composite 3-color plus contour image
of NGC~5253 depicting the optical 6450 \AA\ stellar
continuum (blue), 0.3-8 keV X-rays (green), H$\alpha$ (red)
and total \HI\ column density (contours, as in
Figure~\ref{mom0}).  The X-ray data, previously published in
\citet{summers}, was retrieved from the {\it Chandra X-ray
Observatory} archives (proposal number 02600546) and
adaptively smoothed to a minimum signal-to-noise ratio of 2
for presentation.  Figure~\ref{color} reveals the relative
morphologies of the stars and the cold, warm ionized, and
hot phases of the ISM in NGC~5253.  The X-ray, H$\alpha$,
and \HI\ emission extend preferentially to the south and
west of the nucleus, suggesting expansion of starburst
heated material along these directions. If the \HI\ plume is
inflowing or interacting from the southeast, it may inhibit expansion
in that direction.  

\section{Conclusions}

Radio-wave aperture synthesis observations in the 21-cm line
of neutral hydrogen at 8\arcsec\ resolution have revealed a
complex non-axisymmetric gas distribution and velocity field
in NGC~5253. The total neutral atomic content of the galaxy
is $1.4\times10^8$ \mo\ from single-dish observations for a
distance of 3.8 Mpc.  The neutral medium in NGC~5253 may be
characterized as consisting of three components: 
a component containing the majority of the neutral
medium distributed along the stellar distribution and
showing evidence for low-amplitude rotation at the level of
15--20 \kms;
a redshifted \HI\ plume extending to the south-southeast
along the minor axis containing 20--30\% of the \HI;
and kinematically cold \HI\ shells or
filaments extending along the minor axes and to the
southwest, containing $\sim$10\% of the neutral gas mass.  
Taken together, these different components result in  
very complex velocity structures.  Here we will try to 
summarize out best understanding of these structures, 
starting with the relatively enigmatic \HI\ plume.

\subsection{The Nature of the \HI\ Plume}

The kinematically distinct \HI\ plume consists of a significant fraction
of the total \HI\ emission from NGC~5253.  In this paper we have 
discussed three possible origins for the \HI\ plume: outflowing gas,
inflowing gas, and a separate \HI\ cloud which is interacting with NGC~5253.

The possibility of outflow was motivated by similarity with
velocity fields observed in other dwarf starburst galaxies
showing \HI\ outflow (e.g., NGC 1569, NGC 625, NGC 1705).
However, the large fraction of \HI\ ($\approx$25\%) in the
anamalous velocity gas in NGC~5253 is exceptional.
Furthermore, strong outflows are usually accompanied by
H$\alpha$ and X-ray emission which is not observed at this location.
Such a massive outflow confined to a small solid angle
without corresponding signatures in the ionized gas seems
unlikely.

The possibility of inflow was motivated by the radial velocities of 
CO clouds within the minor-axis dust lane \citep{meier}.     
Our high resolution \HI\ observations
indicate that the plume is a kinematically distinct feature from the \HI\
associated with the molecular gas.  The most compelling argument for
inflow is the near coincidence in velocity of the \HI\
plume with the galaxy's systemic velocity in the region where the two
are spatially coincident.  If the \HI\ plume is infalling, 
it must be from the far side of NGC~5253 given the plume's
velocity field (i.e., the most redshifted gas lies farthest from the galaxy),
consistent with the lack of extinction observed in this vicinity despite
the high \HI\ column density.    

Finally, we have the possibility that the \HI\ plume is a
distinct structure, perhaps a remnant from a recent interaction, seen in
projection against the main body of NGC~5253.  In this
scenario, a recent gravitational encounter, perhaps with
another dwarf galaxy in the same group, produced the
\HI\ plume and provided the trigger for the current extreme
burst of star formation.  Other dwarf galaxies, both
bursting and non-bursting are accompanied by pure \HI\
structures.  Given the relatively large gas mass associated
with the \HI\ plume, this would appear to be the most likely
explanation.

\subsection{A Comprehensive View of the \HI\ in NGC 5253}

Our high resolution view of NGC~5253 has shown us that NGC~5253 has
many similarities to the best studied dwarf starburst galaxies.
\HI\ concentrations near the minor axis dust lane are
coincident in location and velocity with at least two of the
molecular clouds seen in $^{12}$CO \citep{meier}.  The \HI\
extension to the south-southeast, designated here the \HI\
plume, appears to be spatially and kinematically distinct
from the gaseous medium surrounding the dust lane molecular
clouds along the southeast minor axis.  
We present evidence that the redshifted \HI\ plume
is likely to be either a dynamically distinct tidal remnant 
or an inflow on the far side of the galaxy.
\HI\ extensions to the east, west, and southwest 
show remarkably similar morphologies as the
H$\alpha$ shells and filaments.  In particular, there is a
close morphological similarity between the ``quiescent''
H$\alpha$ shell to the west of the nucleus noted by
\citet{martin} and an 800 pc diameter shell-like feature
seen in \HI.  This may be an example of a starburst-powered
shell stalled by the mass of surrounding neutral gas. 

These complex \HI\ features and kinematic signatures, when
considered together with distributions of the molecular,
ionized, and hot coronal media, are consistent with a
starbursting low-mass galaxy in the initial phases of
generating a galactic wind.  Unlike more massive galaxies
such as M~82 or NGC~3079 which show monolithic,
well-collimated outflows \citep[e.g.,][and references therein]
{sh07,cbv02} several non-co-spatial star formation
events over the last $\sim20$ Myr appear to be required to
explain the multiple outflows observed in NGC~5253
\citep{summers, strickland, caldwell}.  The initiating and
sustaining cause of the recent star formation activity may
be an inflow or interaction of atomic gas along the 
``plume'', identified here for the first time.  As such,
NGC~5253 joins a long list of starburst galaxies where
recent gravitational interactions are implicated as the
trigger for extreme star formation bursts.

\acknowledgments

H.~A.~K. was supported through NRA-00-01-LTSA-052.
E.~D.~S. is grateful for partial support from NASA LTSARP grant NAG5-9221 and from 
the University of Minnesota.  We thank an anonymous referee
for incentive to investigate the  
NGC~5253--M~83 (lack of) connection in more detail.
We thank Crystal Martin for supplying us with the continuum and $H\alpha$
images (way back in 1995),  David Meier for 
helpful exchanges regarding the molecular clouds,
and Daniela Calzetti for providing an electronic version of
the Balmer decrement reddening map. 
This research has made use of NASA's Astrophysics Data System Bibliographic Services.

{\it Facilities:} \facility{VLA ()}, \facility{Parkes ()}

{}

\clearpage

\clearpage

\begin{figure}
\plotone{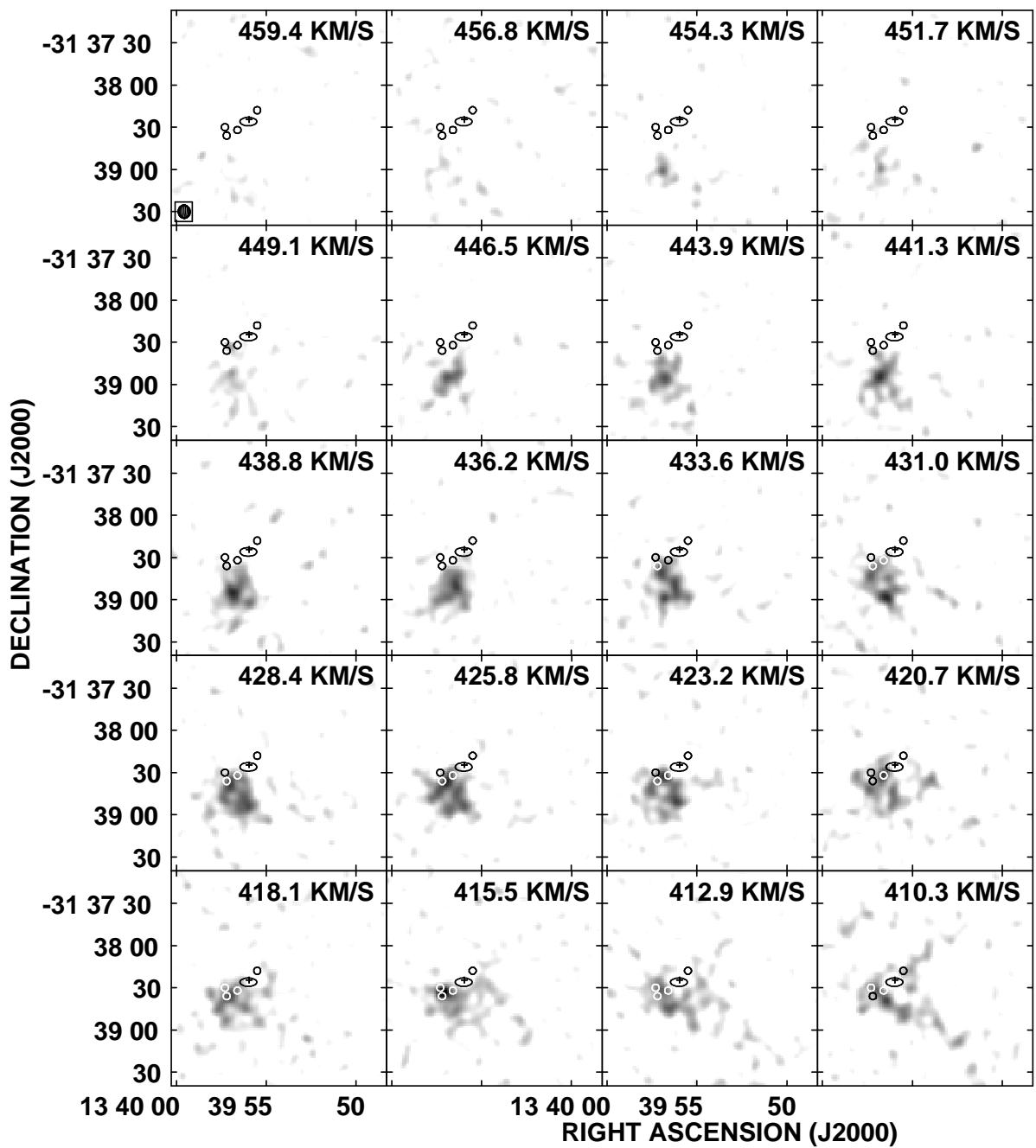}
\caption{\HI\ 21-cm maps of NGC~5253 in every 2.58 \kms\
velocity channel containing emission.  The greyscale
range shows beam-averaged column densities of 1.0$\times10^{20}$ \cmmb\ (white),
equivalent to 2$\times$ the rms noise,
to 5.0$\times10^{20}$ \cmmb\ (black).
The cross marks the location of the central super star cluster and the ellipses are the 
molecular clouds A through E (from left to right) identified by \citet{meier}.
\label{greypanels} }
\end{figure}
\clearpage

\setcounter{figure}{4}

\begin{figure}
\plotone{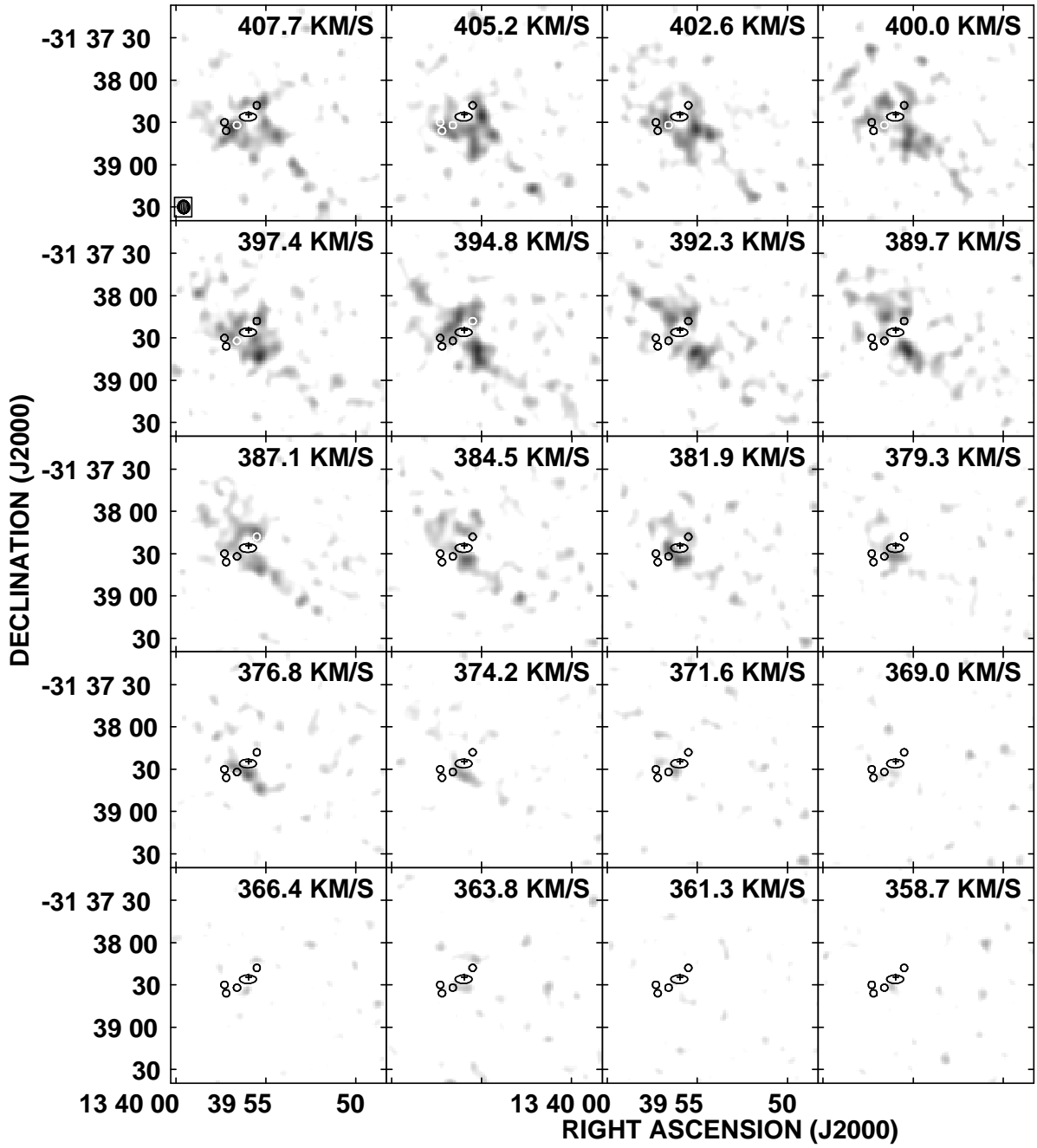}
\caption{continued}
\end{figure}

\clearpage

\begin{figure}
\caption{\HI\ column density channel maps of NGC~5253 (contours)
overlaid on an $H\alpha$ image (greyscale).  
 The contours show fluxes of 2.4, 3.6, 6.0, and 8.4
$\times10^{-3}$ Jy beam$^{-1}$, equivalent to beam-averaged \HI\
column densities of 1.0, 1.5, 2.5, and 3.5 $\times10^{20}$
\cmmb.  \label{panels} }
\end{figure}

\clearpage

\begin{figure}
\plotone{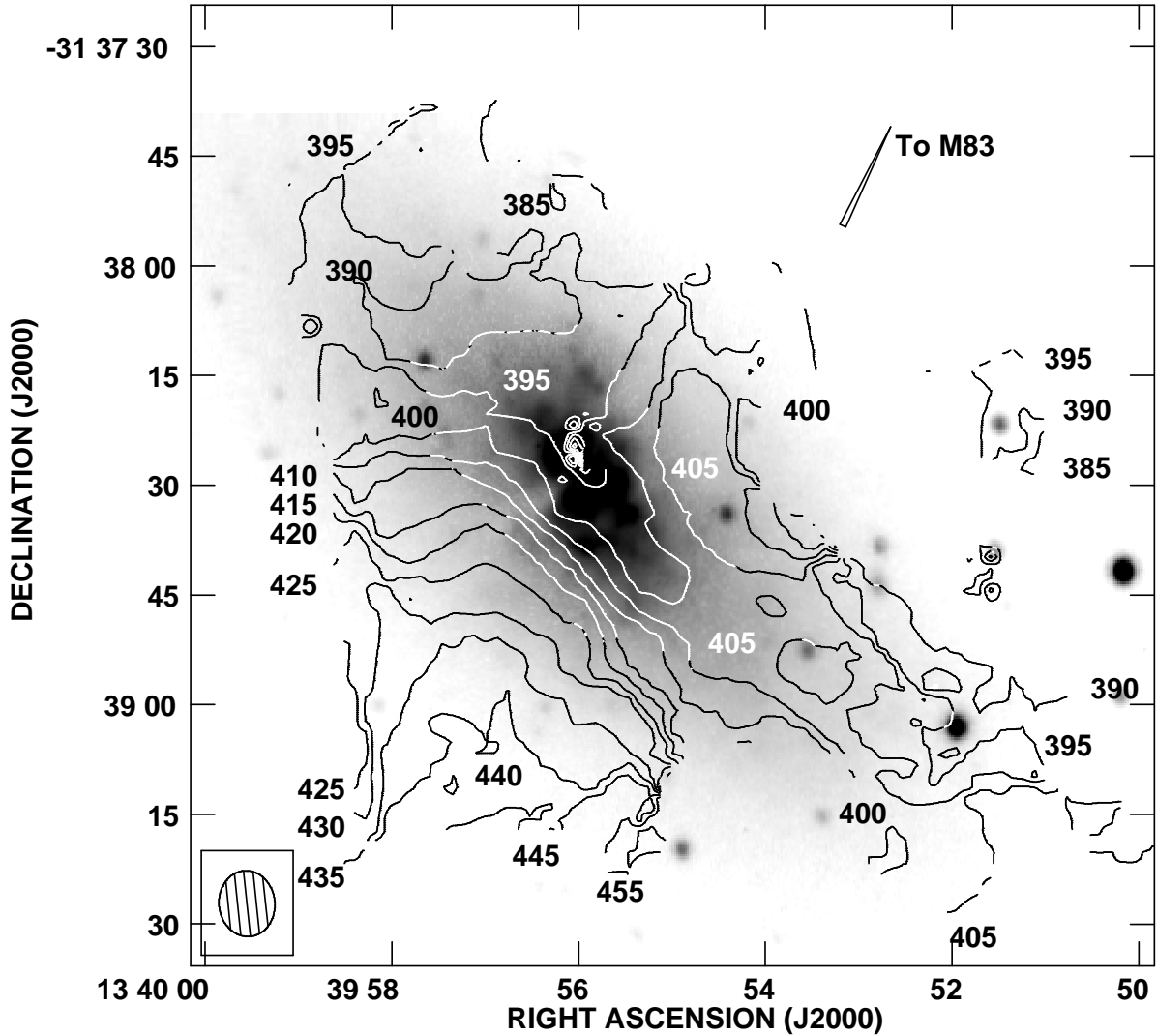}
\caption{A 21-cm intensity-weighted velocity map of NGC~5253  
(i.e., 1st moment; contours) overlaid on an optical 
6450 \AA\ image (greyscale).  Heliocentric isovelocity 
contours are labeled in \kms.  Note the large velocity
gradient in the southeast of the galaxy.  Note also the 
lack of a strong gradient along the major axis of the galaxy.
\label{spider} }
\end{figure}

\clearpage

\begin{figure}
\plotone{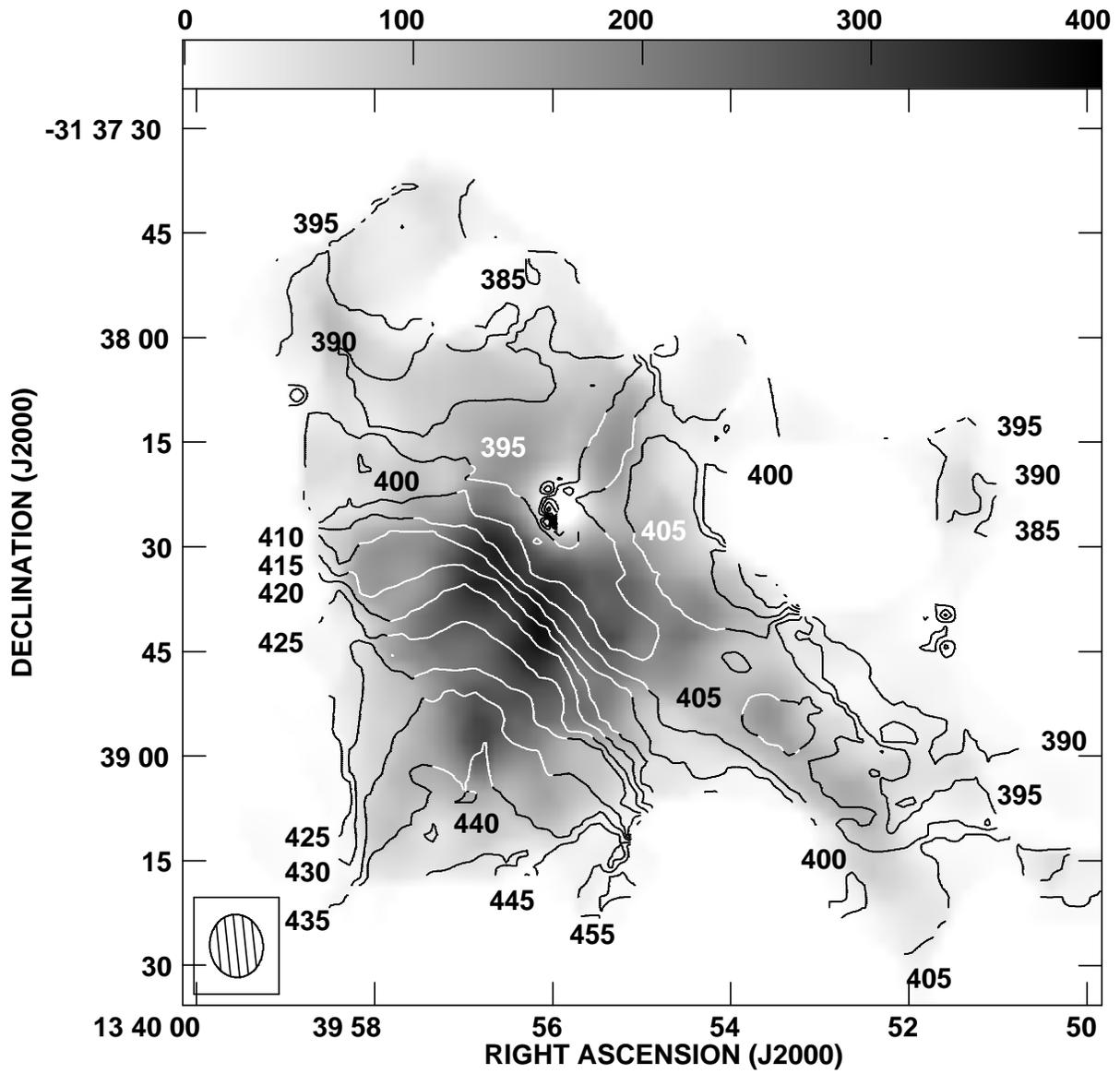}
\caption{A 21-cm intensity-weighted velocity map of NGC~5253  
(i.e., 1st moment; contours) overlaid on a 21-cm total
intensity image (0th moment; greyscale).  Heliocentric isovelocity 
contours are labeled in \kms.  The scale bar shows
\HI\ surface brightness in units of Jy beam$^{-1}$ m s$^{-1}$. 
\label{spider2} }
\end{figure}

\clearpage

\begin{figure}
\plotone{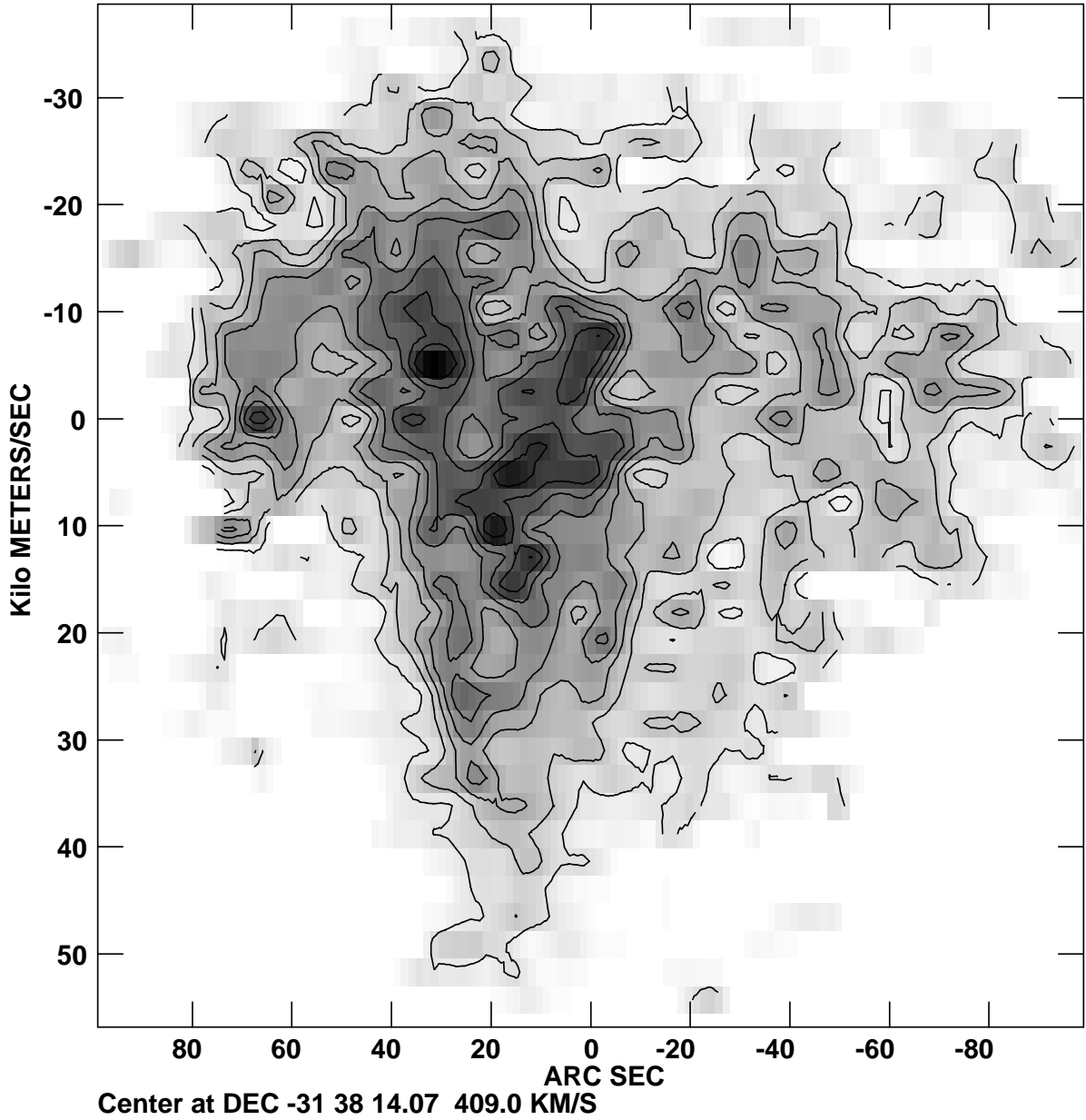}
\caption{A position-velocity diagram taken along the major axis
at position angle 50$^\circ$.  Note the overall trend from left to right
of more positive velocities, and the large velocity range near the center 
of the galaxy.
 \label{lvmajor} }
\end{figure}

\clearpage

\begin{figure}
\plotone{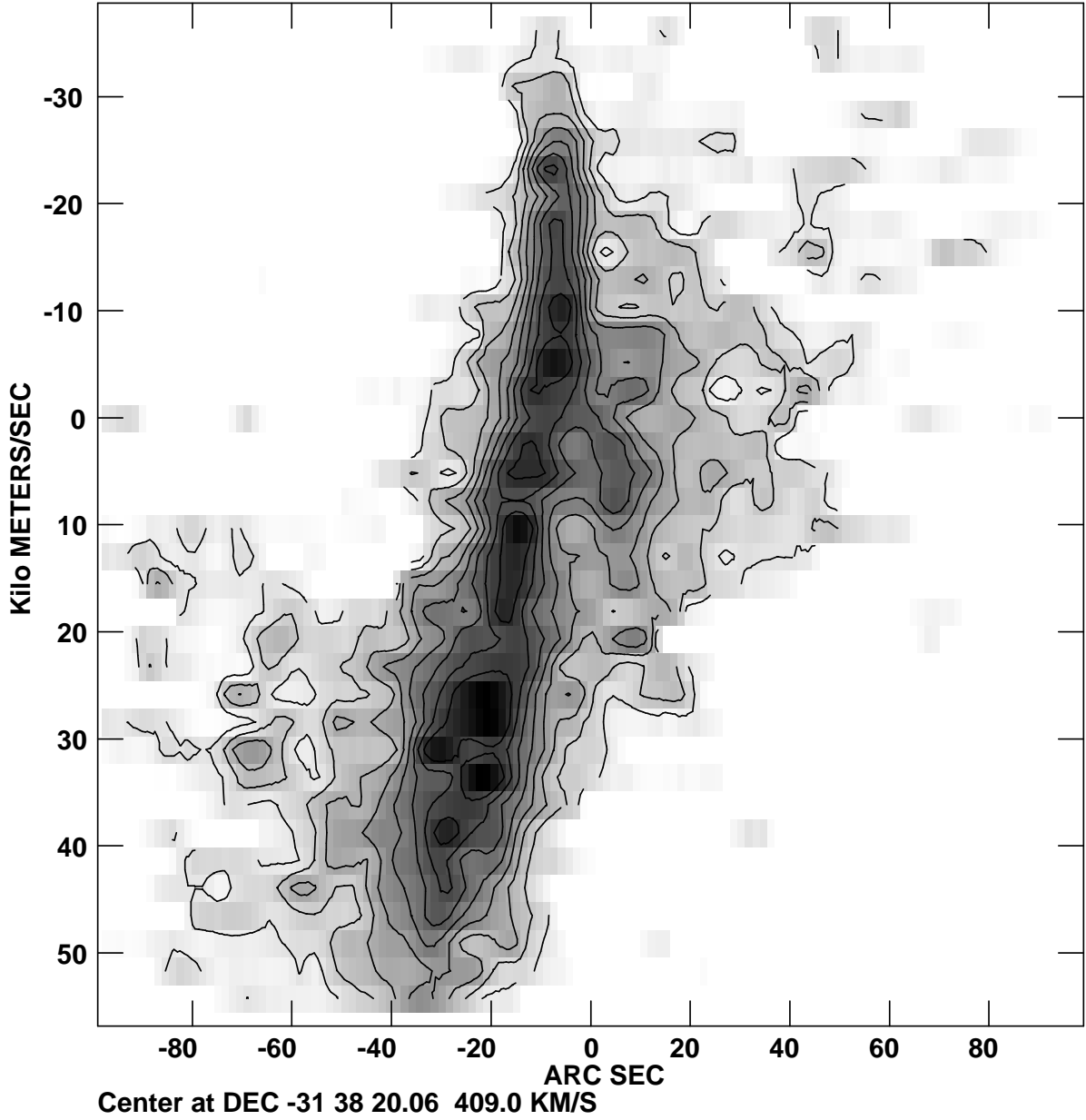}
\caption{A position-velocity diagram taken along the minor axis
at position angle 140$^\circ$.  Note the large, continuous, 
asymmetric velocity range near the center of the galaxy.  
 \label{lvminor} }
\end{figure}

\clearpage

\begin{figure}
\caption{Extinction, $A_V$, as determined from the \HI\ column density (contours)
and the Balmer emission-line decrement (greyscale with colorbar; from \citet{calzetti97}).  
The labeled contours correspond to $A_V(HI)=$4, 3.5, 3, 2.5, 2, 1.5, 1 and 0.5 mag 
assuming a Galactic gas-to-dust ratio. The extended region of
high extinction in the dust lane southeast of the nucleus
has $A_V(H\alpha/H\beta)\simeq1.5$ and is coincident with the second highest peak
in the \HI\ column density at  $A_V(HI)\simeq3.5$.  The \HI\ column density is unreliable in the vicinity of the
central radio supernebula (marked by a cross) owing to \HI\ absorption. 
 \label{AV} }
\end{figure}

\clearpage

\begin{figure}
\caption{A 3-color plus contour view of NGC~5253 showing 
the stars (green), H$\alpha$ (red),
0.3-8 keV X-rays (blue) and total \HI\ column density (contours, as in Figure~\ref{mom0}).
  \label{color} }
\end{figure}

\clearpage


\begin{thebibliography}{99}
\bibitem[Aitken et al.(1982)]{aitken} Aitken, D.~K., Roche, P.~F., 
        Allen, M.~C., \& Phillips, M.~M.  1982, MNRAS, 199, 31P
\bibitem[Alonso-Herrero et al.(2004)]{alonso} Alonso-Herrero, A., Takagi, T., 
	Baker, A., Rieke, G.~H., Rieke, 
	M.~J., Imanishi, M., \& Scoville, N.~Z. 2004, ApJ, 612, 222
\bibitem[Beck et al.(1996)]{beck} Beck, S.~C., Turner, J.~L., Ho, P.~T.~P.,
	Lacy, J.~H., \& Kelly, D.~M. 1996, ApJ, 457, 610
\bibitem[Bloemen et al.(1986)]{bloemen} Bloemen, J.~B.~G.~M., Strong, A.~W., Blitz, L., 
        Cohen, R.~S., Dame, T.~M., Grabelsky, D.~A., Hermsen, W., Lebrun, F., 
        Mayer-Hasselwander, H.~A.,  \& Thaddeus, P. 1986, A\&A, 154, 25
\bibitem[Bohlin et al.(1978)]{bohlin} Bohlin, R.~C., Savage, B.~D., \& Drake, J.~F. 
	1978, ApJ, 224, 132
\bibitem[Caldwell \& Phillips(1989)]{caldwell} Caldwell, N., \& 
	Phillips, M.~M.\ 1989, \apj, 338, 789 
\bibitem[Calzetti et al.(1997)]{calzetti97} Calzetti, D., Meurer, G.~R., 
	Bohlin, R.~C., Garnett, D.~R., 
	Kinney, A.~L., Leitherer, C., \& Storchi-Bergmann, T. 1997, AJ, 114, 1834
\bibitem[Calzetti et al.(1999)]{calzetti99} Calzetti, D., Conselice, C.~J., 
        Gallagher, J.~S. III, \& Kinney, A. 1999, AJ, 118, 797
\bibitem[Calzetti et al.(2004)]{calzetti04} Calzetti, D., Harris, 
	J., Gallagher, J.~S., III, Smith, D.~A., Conselice, C.~J., Homeier, N., \& 
	Kewley, L.\ 2004, \aj, 127, 1405 
\bibitem[Cannon et al.(2003)]{cannon03} Cannon, J.~M., 
	Dohm-Palmer, R.~C., Skillman, E.~D., Bomans, D.~J., C{\^o}t{\'e}, S., \& 
	Miller, B.~W.\ 2003, \aj, 126, 2806 
\bibitem[Cannon et al.(2004)]{cannon04} Cannon, J.~M., McClure-Griffiths, N.~M., 
        Skillman, E.~D., \& C{\^o}t{\'e}, S.\ 2004, \apj, 607, 274
\bibitem[Cannon et al.(2005)]{cannon05} Cannon, J.~M., Skillman, 
	E.~D., Sembach, K.~R., \& Bomans, D.~J.\ 2005, \apj, 618, 247 
\bibitem[Cecil et al.(2002)]{cbv02} Cecil, G., 
	Bland-Hawthorn, J., \& Veilleux, S.\ 2002, \apj, 576, 745 
\bibitem[Clark(1980)]{clark} Clark, B.~G. 1980, A\&A, 89, 377
\bibitem[C{\^o}t{\'e} et al.(2000)]{cote} C{\^o}t{\'e}, S., 
        Carignan, C., \& Freeman, K.~C.\ 2000, \aj, 120, 3027 
\bibitem[Dickey \& Lockman(1990)]{dl} Dicky, J.~M., \& Lockman, F.~J. 1990, ARAA, 28, 215
\bibitem[Draine et al.(2007)]{draine} Draine, B.~T., Dale, D.~A., Bendo, G. 
	et al. 2007, ApJ, in press
\bibitem[Freedman et al.(2001)]{freedman} Freedman, W.,  Madore, B.~F., Gibson, B.~K., 
        Ferrarese, L., Kelson, D.~ D., Sakai, S., Mould, J.~R., Kennicutt, R.~C., Jr., 
	Ford, H.~C., Graham, J.~A., Huchra, J.~P., Hughes, S.~M.~G.,
	Illingworth, G.~D., Macri, L.~M., Stetson, P.~B. 2001, ApJ, 553, 47
\bibitem[Gibson et al.(2000)]{gibson} Gibson, B.~K., et al. 2000, ApJ, 529, 723
\bibitem[Gonzalez-Riestra et al.(1987)]{gonzalez} Gonz\'alez-Riestra, R., 
        Rego, M., \& Zamorano, J. 1987, A\&A, 186, 64
\bibitem[Gorjian et al.(2001)]{gorjian} Gorjian, V., Turner, J.~L., \& 
        Beck, S.~C. 2001, 554, L29
\bibitem[Graham(1979)]{graham79} Graham, J.~A. 1979, ApJ, 232, 60
\bibitem[Graham(1981)]{graham81} Graham, J.~A.\ 1981, \pasp, 93, 552 
\bibitem[Harris et al.(2004)]{harris} Harris, J., Calzetti, D., Gallagher, J.~S.~III, 
        Smith, D.~A., \& Conselice, C.~J. 2004, ApJ, 603, 503
\bibitem[Hodge(1966)]{hodge} Hodge, P.~W.\ 1966, \apj, 146, 593 
\bibitem[{H\"o}gbom(1974)]{hogbom} {H\"ogbom}, J. 1974, ApJS, 15, 417 
\bibitem[Huchtmeier \& Bohnenstengel(1981)]{hucht} Huchtmeier, W.~K., \& 
	Bohnenstengel, H.-D. 1981, AA, 100, 72
\bibitem[Hunter \& Gallagher(1990)]{hunter} Hunter, D.~A. \& Gallagher, J.~S. III, 1990, \apj, 362, 480 
\bibitem[Israel \& Burton(1986)]{israel} Israel, F.~P., \& Burton, W.~B. 1986, A\&A, 168, 369
\bibitem[Kalberla et al.(1982)]{kalberla} Kalberla, P.~W.~M., Mebold, U., \& 
        Reif, K. 1982, A\&A, 106, 190  
\bibitem[Karachentsev et al.(2007)]{kara} Karachentsev, I.~D., et al. 2007, AJ, 133, 504 
\bibitem[Karachentsev et al.(2002)]{karachentsev} Karachentsev, I.~D., Sharina, M.~E., 
        Dolphin, A.~E., Grebel, E.~K., Geisler, D., Guhathakurta, P., Hodge, P.~W.,
	Karachentseva, V.~E., Sarajedini, A., Seitzer, P. 2002, A\&A, 385, 21
\bibitem[Kobulnicky \& Skillman(1995)]{ks} Kobulnicky, H.~A., \& Skillman, 
	E.~D. 1995, ApJ, 454, L121
\bibitem[Kobulnicky \& Johnson(1999)]{kj} Kobulnicky, H.~A., \& 
        Johnson, K.~E. 1999, ApJ, 454, 527, 154
\bibitem[Kobulnicky et al.(1997)]{k97}  Kobulnicky, H.~A.,  Skillman, E.~D., 
         Roy, J--R., Walsh, J., \& Rosa, M.  1997, ApJ,  477, 679
\bibitem[L{\'o}pez-S{\'a}nchez et al.(2007)]{ls07} 
	L{\'o}pez-S{\'a}nchez, {\'A}.~R., Esteban, C., Garc{\'{\i}}a-Rojas, J., 
	Peimbert, M., \& Rodr{\'{\i}}guez, M.\ 2007, \apj, 656, 168 
\bibitem[Mac Low et al.(1989)]{maclow89} Mac Low, M.-M., McCray, R., \& Norman, M.~L.\ 1989, \apj, 337, 141 
\bibitem[Mac Low \& Ferrara(1999)]{maclow} Mac Low, M.-M., \& 
	Ferrara, A.\ 1999, \apj, 513, 142 
\bibitem[Maloney \& Black(1988)]{maloney} Maloney, P., \& Black, J.~H. 1988, ApJ, 325, 389
\bibitem[Marlowe et al.(1995)]{marlowe} Marlowe, A.~T., Heckman, T.~M., 
        Wyse, R.~M., \& Schommer, R. 1995, ApJ, 438, 563
\bibitem[Martin(1997)]{martin97} Martin, C.~L. 1997, ApJ, 491, 561
\bibitem[Martin(1998)]{martin} Martin, C.~L. 1998, ApJ, 506, 222
\bibitem[Martin \& Kennicutt(1995)]{martin95} Martin, C.~L., \& Kennicutt, R.~C., Jr.\ 
         1995, \apj, 447, 171
\bibitem[Mart{\'{\i}}n-Hern{\'a}ndez et al.(2005)]{martin-h} 
	 Mart{\'{\i}}n-Hern{\'a}ndez, N.~L., Schaerer, D., \& Sauvage, M.\ 2005, 
	 \aap, 429, 449 
\bibitem[McMahon et al.(1990)]{mcmahon} McMahon, R.~G., Irwin, M.~J., 
  Giovanelli, R., Haynes, M.~P., Wolfe, A.~M., \& Hazard, C.\ 1990, \apj, 359, 302
\bibitem[Meier et al.(2002)]{meier} Meier, D.~S., Turner, J.~L., \& Beck, S.~C. 
         2002, AJ, 124, 877
\bibitem[Meier et al.(2001)]{meier01} Meier, D.~S., Turner, J.~L., Crosthwaite, L.~P., 
	\& Beck, S.~C. 2001, AJ, 121, 740
\bibitem[Meurer et al.(1998)]{meurer98} Meurer, G.~R., Staveley-Smith, L., \& Killeen,
        N.~E.~B. 1998, \mnras, 300, 705
\bibitem[Mohan et al.(2001)]{mohan} Mohan, N.~R., Anantharamaiah, K.~R., \& 
        Goss, W.~M. 2001, ApJ, 557, 659
\bibitem[Moorwood \& Glass(1982)]{moorwood} Moorwood, A.~F.~M., \& Glass, I.~S. 
         1982, A\&A, 115, 84
\bibitem[Saha et al.(1995)]{saha} Saha, A., Sandage, A., 
         Labhardt, L., Schwengeler, H., Tammann, G.~A., Panagia, N., \& Macchetto, 
         F.~D.\ 1995, \apj, 438, 8 
\bibitem[Sakai et al.(2004)]{sakai} Sakai, S., Ferrarese, L., 
         Kennicutt, R.~C., Jr., \& Saha, A.\ 2004, \apj, 608, 42 
\bibitem[Scoville et al.(1987)]{scoville} Scoville, N.~Z., Yun, M.~S., 
         Clemens, D.~P., Sanders, D.~B., \& Waller, W.~H. 1987, ApJS, 63, 821
\bibitem[Shostak \& Skillman(1989)]{ss89} Shostak, G.~S., \& 
	 Skillman, E.~D.\ 1989, \aap, 214, 33 
\bibitem[Stil \& Israel(2002)]{stil} Stil, J.~M., \& Israel, 
	 F.~P.\ 2002, \aap, 392, 473
\bibitem[Strickland \& Heckman(2007)]{sh07} Strickland, 
	 D.~K., \& Heckman, T.~M.\ 2007, \apj, 658, 258 
\bibitem[Strickland \& Stevens(1999)]{strickland} Strickland, D.~K., \& 
         Stevens, I.~R. 1999, MNRAS, 306, 43
\bibitem[Strickland et al.(2004)]{strickland04} Strickland, D.~K., Heckman, T.~M., 
         Colbert, E.~J.~M., Hoopes, C.~G., \& Weaver, K.~A.\ 2004, \apj, 606, 829 
\bibitem[Summers et al.(2004)]{summers} Summers, L.~K., Stevens, I.~R., 
         Strickland, D.~K., \& Heckman, T.~M. 2004, MNRAS, 351, 1 
\bibitem[Taylor(1997)]{taylor} Taylor, C.~L.\ 1997, \apj, 480, 524
\bibitem[Thim et al.(2003)]{thim} Thim, F., Tammann, G.~A., Saha, A., Dolphin, A., 
	Sandage, A., Tolstoy, E., Labhardt, L. 2003, ApJ, 590, 526
\bibitem[Tremonti et al.(2001)]{tremonti} Tremonti, C.~A., Calzetti, D., 
        Leitherer, C., \& Heckman, T.~M.  2001, ApJ, 555, 322
\bibitem[Turner, Beck, \& Ho(2000)]{turner00} Turner, J.~L., Beck, S.~C., \& 
        Ho, P.~T.~P. 2000, ApJ, 532, L109
\bibitem[Turner et al.(1998)]{turner98} Turner, J.~L., Ho, P.~T.~P., \& 
        Beck, S.~C. 1998, AJ, 116, 1212
\bibitem[Turner, Beck, \& Hurt(1997)]{turner97} Turner, J.~L., Beck, S.~C., \& Hurt, R.~L. 
	1997, ApJ, 474, L11
\bibitem[Ulvestad et al.(2006)]{ulvestad} Ulvestad, J.~S, Perley, R.~A., \& Taylor, G.~B. 2006, 
	The VLA Observational Status Summary, 
\bibitem[van den Bergh(1980)]{vdbergh} van den Bergh, S.\ 1980, \pasp, 92, 122 
\bibitem[Vanzi \& Sauvage(2004)]{vanzi} Vanzi, L., \& Sauvage, M. 2004, A\&A, 415, 509
\bibitem[Verter \& Hodge(1995)]{verter} Verter, F., \& Hodge, P. 1995
\bibitem[Walsh \& Roy(1989)]{wr} Walsh, J.~E., \& Roy, J-R. 1989, MNRAS, 239, 297
\bibitem[Wilcots \& Miller(1998)]{wm98} Wilcots, E.~M., \& 
         Miller, B.~W.\ 1998, \aj, 116, 2363 
\bibitem[Wood \& Churchwell(1989)]{wood} Wood, D.~O., \& Churchwell, E.~B. 1989, ApJS, 69, 831
\end{thebibliography}
\end{document}